\documentclass[aps,11pt,english,tightenlinesletterpaper,superscriptaddress,secnumarabic,nofootinbib]{revtex4}

\pdfoutput=1

\usepackage{multirow}
\usepackage{mypackages}
\usepackage[sort&compress]{natbib}
\usepackage{hyperref} 	
\usepackage[all]{hypcap}  
\hypersetup{
    colorlinks=true,		
    linkcolor=blue,		
    citecolor=green,		
    filecolor=magenta,	
    urlcolor=cyan		
}

\usepackage{myterms}
\newcommand{\LamLW}{\Lambda_{\text{\sc lw}}}
\newcommand{\LamH}{\Lambda_{H}}
\newcommand{\LamW}{\Lambda_{W}}
\newcommand{\LamB}{\Lambda_{B}}
\newcommand{\LamZ}{\Lambda_{Z}}
\newcommand{\LamA}{\Lambda_{A}}
\newcommand{\LamAZ}{\Lambda_{AZ}}
\newcommand{\LamEW}{\Lambda_{\rm EW}}
\newcommand{\Lamt}{\Lambda_{t}}
\newcommand{\LamQ}{\Lambda_{Q}}
\newcommand{\Lamu}{\Lambda_{u}}
\newcommand{\gast}{g_{\ast}}

\begin{document}

\title{The Lee-Wick Standard Model at Finite Temperature}

\author{
Richard F. Lebed, Andrew J. Long, and Russell H. TerBeek \\ 
Department of Physics, Arizona State University, Tempe, AZ 85287-1504 \\ 
\ E-mail:\ richard.lebed@asu.edu, andrewjlong@asu.edu, russell.terbeek@asu.edu
}

\begin{abstract}
The Lee-Wick Standard Model at temperatures near the
  electroweak scale is considered, with the aim of studying the
  electroweak phase transition.
  While Lee-Wick theories possess states of negative norm, they are
  not pathological but instead are treated by imposing particular
  boundary conditions and using particular integration contours in the
  calculation of S-matrix elements.  It is not immediately clear how
  to extend this prescription to formulate the theory at finite
  temperature; we explore two different pictures of finite-temperature
  LW theories, and calculate the thermodynamic variables and the
  (one-loop) thermal effective potential.  We apply these results to
  study the Lee-Wick Standard Model and find that the electroweak
  phase transition is a continuous crossover, much like in the
  Standard Model.  However, the high-temperature behavior is modified
  due to cancellations between thermal corrections arising from the
  negative- and positive-norm states.  
\end{abstract}


\maketitle

\setlength{\parindent}{20pt}
\setlength{\parskip}{2.5ex}

\section{Introduction}\label{sec:Introduction}

The Lee-Wick Standard Model (LWSM) \cite{Grinstein:2007mp} is an
extension of the Standard Model (SM) that tames the Higgs mass
hierarchy problem by modifying the dispersion relationships of the
various SM fields in order to improve the UV behavior of the theory.
This modification is accomplished by introducing a new mass scale
$\LamLW$ and extending the Lagrangian by dimension-six operators of
the forms $\delta \mathcal{L} = (\Box \phi)^2 / \LamLW^2$ and $\delta
\mathcal{L} = \bar{\Psi} i (\slashed{\partial})^3 \Psi / \LamLW^2$ and
$\delta \mathcal{L} = {\rm Tr}\left[ D^{\mu} F_{\mu \nu} D^{\alpha}
  F_{\alpha \beta} \right] g^{\nu \beta} / \LamLW^2$.  As such, the
propagators fall off more rapidly in the ultraviolet (UV) limit above
the scale $\LamLW$, which softens the divergences in one-loop
corrections to the Higgs self-energy from dangerous quadratic ones to
harmless logarithmic ones.  To eliminate the need for fine tuning,
$\LamLW$ should be not much larger than the electroweak scale.

Since the LWSM augments the SM by new degrees of freedom at the
electroweak (EW) scale that are coupled to the Higgs (and indeed, one
can study a variant LWSM in which only these fields are
significant~\cite{Carone:2008bs}), it is natural to expect the new
physics to affect the nature of the electroweak phase transition.
This connection is further motivated by the relationship between UV
quadratic divergences and the phenomenon of symmetry restoration
\cite{Comelli:1996vm}.  That is, the same Feynman graphs that give
rise to quadratic divergences in the Higgs self-energy also yield
$O(T^2)$ corrections to the effective mass at finite temperature, and
thereby lift the tachyonic Higgs mass and induce symmetry restoration.
In previous work, the free energy density and thermodynamic properties
of the LWSM plasma have been calculated
\cite{Fornal:2009xc}, and the non-thermal effective potential has been
derived \cite{Espinosa:2011js}.  The goal of this paper is to study
the LWSM at finite temperature using the thermal effective potential
in order to determine the nature of the electroweak phase transition
and symmetry restoration.

The LWSM Lagrangian contains higher-order time derivatives of the
various SM fields, which leads to roughly a doubling of the number of
dynamical degrees of freedom.\footnote{Actually, the new vector
degrees of freedom are massive, and Dirac fermion partners pick up
extra poles, and therefore the number of degrees of freedom is
somewhat more than doubled.  This point is discussed in
\sref{sub:LWSM_Veff}; it plays an important role in the issue of
symmetry restoration.  }  It is well-known that such higher-derivative
(HD) theories generally suffer from a variety of pathologies (see,
{\it e.g.},~\cite{Hawking:1985gh,Hawking:2001yt} for a pedagogical
discussion).  At the classical level, Ostrogradsky's theorem forces
the Hamiltonian to be unbounded from below due to excitations of the
new degrees of freedom.  If one departs from the canonical
quantization prescription in quantizing the theory, then the spectrum
can be rendered bounded from below, but at the cost of introducing
states of negative norm, {\it i.e.}, ghosts.  Lee and Wick developed a
prescription for removing the ghosts and rendering the theory
predictive by treating the system as a boundary-value problem and
imposing boundary conditions at future infinity \cite{Lee:1969fy,
Lee:1970iw} (see also
\cite{Cutkosky:1969fq}).  Subject to these boundary conditions, the
system develops an acausal behavior on the timescale $\LamLW^{-1}$.
For $\LamLW = O(\rm{TeV})$, the acausality is confined to microscopic
scales, and thereby evades constraints from direct laboratory
observation.

Due to the pathologies of HD theories, it is {\it a priori} unclear
how to correctly and consistently formulate a calculation at finite
temperature.  
To illustrate where the trouble arises, consider a
classical HD theory.  At finite temperature, a system approaches
thermal equilibrium by redistributing energy between its many degrees
of freedom so as to minimize its energy and maximize its entropy.
However, for a system in which the Hamiltonian is unbounded, the
entropy can always be increased without bound by lowering the energy
of some degrees of freedom and raising the energy of others.  This
discussion illustrates why care must be taken in formulating the
calculation at finite temperature.

Developing the correct formulation of Lee-Wick theories at finite temperature is one of the goals of this paper.  
Previous efforts to tackle this problem have taken different approaches: One group studied an ideal gas of negative-energy particles \cite{Bhattacharya:2011bb}, whereas a second group studied the partition function of SM particles that are able to scatter through negative-norm narrow resonances \cite{Fornal:2009xc}.  
Both groups concluded that the contribution from a LW field to the free energy is precisely the opposite of that from a SM field with identical mass and spin.  
However, a third group argued that the connection between symmetry restoration and UV behavior suggests that a relative minus sign should not appear \cite{Espinosa:2011js}.  
In \sref{sec:Thermo_of_LW} we explore and extend the previous work by first addressing whether it is more realistic to treat the LW fields as an ideal gas or as resonances, and second by addressing the issue of the sign.  
On the second point, we introduce an index $\sigma = \pm 1$ in order to consider both sign choices simultaneously and thereby keep our analysis general.  
For the case $\sigma = +1$($-1$), the LW fields contribute to the free energy density with the same (opposite) sign as SM fields, and it is from this perspective that we proceed to study the LWSM at finite temperature.  
We find that the two cases lead to qualitatively different outcomes with regard to the temperature of the electroweak phase transition, as well as the sign of thermodynamic quantities in the ultra-relativistic limit.  

This paper is organized as follows.  For the reader who is unfamiliar
with Lee-Wick theories, we provide a more detailed introduction to the
subject in \sref{sec:Intro_to_LW}.  In \sref{sec:Thermo_of_LW} we
formulate the thermodynamics of Lee-Wick theories and calculate the
one-loop thermal effective potential for a toy model.  In
\sref{sec:LWSM_finiteT} we evaluate the thermal effective potential
for the LWSM and study the LWSM at finite temperature, determine the
nature of the electroweak phase transition, and investigate the
phenomenon of symmetry restoration.  In \sref{sec:Conclusions} we
summarize and conclude.  Appendix~\ref{sec:quant_conv} describes
possible quantization conventions, and App.~\ref{sec:Derive_Masses}
gives details of the LWSM spectrum.

\section{Introduction to Lee-Wick Theories}\label{sec:Intro_to_LW}

The Lee-Wick Standard Model~\cite{Grinstein:2007mp} was developed by
Grinstein, O'Connell, and Wise as an alternative approach to taming
the gauge hierarchy problem of the Standard Model.  In the case of the
much better-explored example of low-scale supersymmetry (SUSY), each
SM loop diagram is joined by one in which the loop particle is
replaced by its opposite-statistics partner (but which carries the
same gauge and Yukawa couplings), thus introducing a relative sign
difference that induces the cancellation of the leading-order
(quadratic) divergence.  In the LWSM, each loop diagram is joined by
one in which the loop particle is replaced by its opposite-{\it
norm\/} partner, again inducing the desired cancellation.

The essence of the original Lee and Wick program~\cite{Lee:1969fy,
  Lee:1970iw} is the promotion of Pauli-Villars regulators to the
status of full dynamical fields with negative quantum-mechanical norm.
Obviously, such unusual states introduce paradoxes of physical
interpretation that must be addressed.  At the classical level, such
signs correspond to instabilities in the form of runaway states of
ever-increasing negative energy, while at the quantum level a negative
norm (which generates a Hilbert space of indefinite
metric~\cite{Pauli:1943xx}) produces a violation of unitarity.
However, Lee and Wick showed that these runaway solutions can be
eliminated from the theory by the imposition of future boundary
conditions on Green's functions, which has the price of introducing
violations of causality.  If the LW scale is sufficiently high, then
the realm of acausal effects is relegated to an unobservably
microscopic scale.  Moreover, if the negative-norm states are required
to be unstable (decaying into conventional particles), then they may
be excluded from the set of asymptotic states of the theory, thus
restoring unitarity.  In order for the exclusion of on-shell
negative-norm states to make sense in Feynman loop diagrams, Lee and
Wick developed a variant of the Feynman integration contour for such
cases, a program that was greatly expanded by Cutkosky {\it et
  al.}~\cite{Cutkosky:1969fq} (CLOP).  While no problematic exceptions
to this program are known, it remains unknown whether a
nonperturbative formulation exists that preserves
unitarity~\cite{Boulware:1983vw}.

In the same way that adding a Pauli-Villars regulator to a scalar
propagator softens its high-momentum behavior from $~1/p^2$ to
$~1/p^4$, the Lagrangian of a scalar theory containing a particle and
its LW partner is promoted from one with a canonical $\partial^2$
kinetic energy term to a higher-derivative theory with a
$\partial^4$ term.  To be explicit, let $\hat \phi$ be a real scalar
field appearing in the Lagrangian
\begin{equation}
{\cal L}_{{\rm HD}}=-\frac{1}{2} \hat{\phi} \,\Box\, \hat{\phi}
- \frac{1}{2\LamLW^2} \hat{\phi} \,\Box^2 \hat{\phi}
-\frac{1}{2} m^2 \hat{\phi}^2 +{\cal L}_{{\rm int}}(\hat{\phi})
\, ,
\label{eq:toyhd}
\end{equation}
where the last term represents interactions.  One may recast
Eq.~(\ref{eq:toyhd}) in an equivalent form without the HD term by
introducing an auxiliary field (AF) $\tilde{\phi}$:
\begin{equation}
{\cal L}_{{\rm AF}}= -\frac{1}{2} \hat{\phi} \,\Box\, \hat{\phi}
-\frac{1}{2} m^2 \hat{\phi}^2 - \tilde{\phi} \,\Box\, \hat{\phi} +
\frac{1}{2} \LamLW^2 \tilde{\phi}^2 +{\cal L}_{{\rm
int}}(\hat{\phi})
\, .
\label{eq:toyaf}
\end{equation}
The equation of motion for $\tilde{\phi}$,
\begin{equation}
\tilde{\phi} = \frac{1}{\LamLW^2} \Box\,\hat{\phi} \, ,
\end{equation}
is exact at the quantum level (meaning that the path integral over
this degree of freedom can be performed exactly), and upon
substitution into Eq.~(\ref{eq:toyaf}), reproduces
Eq.~(\ref{eq:toyhd}).  Further defining the field $\phi \equiv
\hat{\phi} + \tilde{\phi}$ diagonalizes the kinetic energy terms:
\begin{equation}
{\cal L}=-\frac{1}{2} \phi \,\Box\, \phi+\frac{1}{2} \tilde{\phi}
\,\Box\, \tilde{\phi} -\frac{1}{2} m^2
(\phi-\tilde{\phi})^2+\frac{1}{2} \LamLW^2 \tilde{\phi}^2
+{\cal L}_{{\rm int}}(\phi-\tilde{\phi}) \, .
\end{equation}
One diagonalizes the mixed mass terms without altering the kinetic
terms by a symplectic transformation:
\begin{equation}
\left(\begin{array}{c} \phi \\ \tilde{\phi}\end{array}\right) =
\left(\begin{array}{cc} \cosh\theta & \sinh\theta \\ \sinh\theta &
\cosh\theta \end{array}\right) \left(\begin{array}{c} \phi_0 \\
\tilde{\phi}_0 \end{array}
\right) \, ,
\end{equation}
with mass eigenstates being indicated by subscript $0$, and the
transformation parameter $\theta$ satisfies
\begin{equation}
\tanh 2\theta = \frac{-2 m^2}{\LamLW^2 -2 m^2}  \, ,
\end{equation}
which admits real solutions provided $\LamLW^2 > 4m^2$.  If this LW
stability condition fails, then the kinetic and mass terms cannot be
simultaneously diagonalized with real mass eigenvalues, and the
Lagrangian \eref{eq:toyhd} does not represent a Lee-Wick theory.  The
Lagrangian then assumes the form
\begin{equation}
{\cal L}_{{\rm LW}} = -\frac{1}{2} \phi_0 \,\Box\, \phi_0+\frac{1}{2}
\tilde{\phi}_0 \,\Box\, \tilde{\phi}_0 -\frac{1}{2} m_0^2 \phi_0^2
+\frac{1}{2} M_0^2 \tilde{\phi}_0^2 +{\cal L}_{{\rm
int}}[e^{-\theta}(\phi_0-\tilde{\phi}_0)] \, ,
\label{eq:toylw}
\end{equation}
for mass eigenvalues
\begin{equation}
  m_0^2, \, M_0^2 \equiv \frac{\LamLW^2}{2} \left( 1 \mp
\sqrt{ 1 - \frac{4m^2}{\LamLW^2}} \right) \, ,
\end{equation}
and the factor of $e^{-\theta}$ can be absorbed into redefinitions of
the couplings.  The quadratic terms in Eq.~(\ref{eq:toylw}) clearly
manifest the promised opposite-norm $\phi_0$ and $\tilde{\phi}_0$
propagators (see App.~\ref{sec:quant_conv}).  
This fact, combined with the fixed relationship between
$\phi_0$ and $\tilde{\phi}_0$ couplings seen in ${\cal L}_{{\rm
    int}}$, leads to the cancellation of quadratic divergences, as
shown explicitly in Ref.~\cite{Grinstein:2007mp}.  While we have
presented only the LW construction for a real scalar field, an
analogous AF construction holds for all SM
fields~\cite{Grinstein:2007mp}: complex scalars (with or without
spontaneous symmetry breaking), Dirac fermions, and vector fields
(including gauge fields).

We see that SM particles with LW partners can be represented by HD
fields appearing in a restricted class of Lagrangians (so that the
mass eigenvalues turn out real and positive) whose propagators fall
off as $1/p^4$ and have two propagator poles, which represent one
field of positive and one of negative norm.  But nothing in principle
requires the HD theory to truncate at just two extra derivatives.  One
can define a LW theory of a given $N$ as one in which the full
propagator has $N$ poles, or equivalently, $2N$ extra derivatives in
the Lagrangian.  The SM would therefore be called an $N=1$ theory, the
LWSM would be $N=2$, and as shown in Ref.~\cite{Carone:2008iw}, one
can build $N \ge 3$ theories for all fields appearing the SM,
including a proper AF construction.  Furthermore, one finds that the
additional field degrees of freedom alternate in norm: Each $N=3$
field, like its SM partner, has positive norm.  Such a generalized LW
theory is quite unlike SUSY and rather more resembles theories with
Kaluza-Klein (KK) excitations, such as extra-dimension models.

Nevertheless, LW theories are unlike both SUSY and KK theories in
important respects.  Since no principle dictates how many LW partners
a given SM field possesses nor what determines the LW scale, one can
imagine a scenario in which some SM fields have 2 partners, some have
1, and some have none.  In contrast, the closure of the SUSY algebra
requires every field to have precisely one opposite-statistics
partner, while KK theories have no predetermined limit on the number
of modes available to the field.  This generality of LW theories of
course comes at a price.  To name just a few issues: In fits to data or in
making predictions, one must allow for the possibility that all field
LW mass scales are distinct; the equivalent HD theory may only be an
effective theory of an unknown UV completion (for our purposes, we
assume only that the effective theory is good up to the 14~TeV reach
of the Large Hadron Collider); and while grand unification is
possible~\cite{Grinstein:2008qq,Carone:2009it}, it is not as
straightforward to arrange as in, say, the MSSM\@.  Even so, LW
theories are quite flexible and can be combined with other beyond-SM
(BSM) ideas like SUSY~\cite{Dias:2012fi,Gama:2011ws}.

The LWSM was subjected to tests of its phenomenological viability as a
potential BSM theory already starting in Ref.~\cite{Grinstein:2007mp},
and subsequently compared to precision electroweak constraints in a
variety of interesting
ways~\cite{Alvarez:2008za,Underwood:2008cr,Carone:2008bs,
Chivukula:2010nw,Krauss:2007bz,Carone:2009nu,Alvarez:2011ah,
Figy:2011yu}.  The consensus view emerged that LW gauge bosons must
have masses at least $~2$~TeV and the LW fermions at least several
TeV, but the LW scalars can be substantially lighter.  When $N=3$
partners are permitted, the allowed gauge boson partner masses must
still be at least 2~TeV or higher, and the fermions may be as low as
$1.5$~TeV, but viable scenarios in which the scalar partners lie in
the several hundred GeV range emerge~\cite{Lebed:2012ab}.

\section{Thermodynamics of Lee-Wick Theories}\label{sec:Thermo_of_LW}

In this section we address the question of how one should calculate
the thermodynamic properties ({\it e.g.}, entropy, energy density) of
a LW theory.  It is unclear to what extent the standard formulation of
this calculation is applicable due to the presence of unphysical
degrees of freedom, namely, the negative-norm LW particles.  At zero
temperature, one imposes boundary conditions to remove the LW
particles from the set of asymptotic states and employs the LW/CLOP
prescriptions to calculate elements of the unitary S~matrix between
states containing only SM particles.  It is not obvious how to extend
the boundary conditions and LW / CLOP prescriptions to a LW theory at
finite temperature.  Thus, two pictures emerge: Either
\begin{itemize}
\item The thermal
system can access states containing explicit LW particles, or
\item The system can only explore states from which these explicit LW particles
are absent.  
\end{itemize}
Both scenarios have been considered in the literature
(\cite{Bhattacharya:2011bb} and \cite{Fornal:2009xc}, respectively).  In fact, Ref. \cite{Bhattacharya:2011bb} obtains the same result for the free energy as Ref. \cite{Fornal:2009xc}.  We argue, however, that
the pictures are not equivalent, but instead that the second picture,
in which LW particles only serve to modify the scattering of SM
particles, is more realistic.  In the next subsection we show that no self-consistent calculation using ideal gas LW particles appears to agree with the common result of Refs. \cite{Fornal:2009xc, Bhattacharya:2011bb}.  Furthermore, Ref. \cite{Bhattacharya:2011bb} uses a convention of negative-energy, positive-norm particles, while Ref. \cite{Fornal:2009xc} uses negative-norm, positive-energy particles.  While we adopt the second convention, the first one can be shown to be equivalent if properly implemented (see App. A).

\subsection{Ideal Gas of LW Particles}\label{sub:IdealGasLWParticles}

In this section, we consider the first of the two pictures discussed
above and calculate the thermodynamic properties of an ideal gas of LW
particles.  A LW theory contains both SM and LW particles, but in the
absence of interactions, their ideal gas contributions can be
evaluated separately.  We define the partition function $Z$ by the
requirement that the density matrix,
\begin{align}\label{eq:def_densitymatrix}
	\hat{\rho} = \frac{1}{Z} {\rm exp}( - \beta \hat{H}) \, ,
\end{align}
is properly normalized (see below).  With interactions turned off, the
spectrum of the Hamiltonian $\hat{H}$ consists of the vacuum
$\ket{0}$, single-particle states $\ket{\bf p}$, and multi-particle
states $\ket{{\bf p}_1, {\bf p}_2, \ldots, {\bf p}_N}$ with the
appropriate symmetrization (anti-symmetrization) for bosons
(fermions).  For example,
\begin{align}
	\ket{{\bf p}_1 \,  {\bf p}_2} = \frac{1}{\sqrt{2}}
\left( \ket{{\bf p}_1} \otimes \ket{{\bf p}_2} + \eta_S \ket{{\bf p}_2}
\otimes \ket{{\bf p}_1} \right) \, ,
\end{align}
where $\eta_S = +1$ ($-1$) for bosons (fermions).  The single-particle
states satisfy $\hat{H} \ket{{\bf p}} = E_{\bf p} \ket{\bf p}$, where
$E_{\bf p} = \sqrt{ {\bf p}^2 + m^2 }$.  These expressions use the
quantization convention $\eta_C = +1$ of Eq.~(\ref{canonicalcom2}).

As discussed in \sref{sec:Intro_to_LW}, states with an odd number of
LW particles have a negative norm due to the wrong-sign commutation
relation of the associated creation and annihilation operators.  We
use the index $\eta_N$ [see Eq.~(\ref{canonicalcom2})] to keep track of
this norm; for LW particles (SM particles) we have $\eta_N=-1$ ($+1$).
For example,
\begin{align}
\begin{array}{l}
	\amp{0}{0} =  1 \, , \\
	\amp{{\bf p}}{{\bf q}} =  \eta_N (2\pi)^3 \, 2 E_{\bf p} \,
\delta({\bf p}-{\bf q}) \, , \\
	\amp{{\bf p}_1 \, {\bf p}_2 }{{\bf q}_1 \, {\bf q}_2} =
(2\pi)^6 \, 2E_{{\bf p}_1} 2E_{{\bf p}_2} \left[
\delta({\bf p}_1-{\bf q}_1) \, \delta({\bf p}_2-{\bf q}_2)
	+ \eta_S \, \delta({\bf p}_1-{\bf q}_2) \,
\delta({\bf p}_2-{\bf q}_1) \right] \, ,
\end{array}
\end{align}
and so on.  The negative norm implies that eigenvalues and expectation
values differ by a sign.  For instance,
\begin{align}
	\int \frac{d^3 q}{(2\pi)^3} \frac{1}{2 E_{\bf q}}
\expval{{\bf p}}{\hat{H}}{{\bf q}} = \eta_N E_{\bf p} \com
\end{align}
whereas the state $\ket{\bf p}$ has eigenvalue $E_{\bf p}$.  This
distinction is particularly relevant for the calculation of the
partition function.  If we normalize the density matrix by requiring
\begin{align}\label{eq:rho_norm_1}
	{\rm Tr} \, \hat{\rho} = 1 \com
\end{align}
then the partition function $Z = {\rm Tr} \, e^{- \beta \hat{H}}$ is
given by a sum of expectation values
\begin{align}
	Z 
	= \expval{0}{e^{-\beta \hat{H}}}{0} 
	+ \int \frac{d^3 p}{(2\pi)^3} \frac{1}{2 E_{\bf p}} \expval{{\bf p}}{e^{- \beta \hat{H}}}{{\bf p}} 
	+ \int \frac{d^3 p}{(2\pi)^3} \frac{1}{2 E_{\bf p}} \int \frac{d^3 q}{(2\pi)^3} \frac{1}{2 E_{\bf q}} \expval{{\bf p}, {\bf q}}{e^{- \beta \hat{H}}}{{\bf p}, {\bf q}} + \ldots \per
\end{align}
In the case $\eta_N = -1$, the terms alternate in sign.  Since the
expectation values of $\hat{\rho}$ are not strictly positive, the
possibility may arise that the sum of the eigenvalues of $\hat{\rho}$
becomes greater than unity, while $\hat{\rho}$ itself remains
normalized in the sense of \eref{eq:rho_norm_1}.  It is not clear how
to interpret such a density matrix.  Alternatively, one can normalize
the density matrix by imposing
\begin{align}\label{eq:rho_norm_2}
	{\rm Tr}^{\prime} \hat{\rho} \equiv \sum_{\rm eigs}
\hat{\rho} = 1 \, ,
\end{align}
where Trace$^\prime$ is obtained by simply summing the eigenvalue
spectrum of the operator.  In this case, the norm of the states is
irrelevant to the calculation, and its outcome is the standard ideal
gas partition function.  It is not {\it a priori\/} clear that this sum is finite; such an assertion is equivalent to assuming that the conditionally convergent series implied in Eqn.\,\eqref{eq:rho_norm_1} is absolutely convergent. We do not dwell on the issue of which
normalization condition is the ``correct'' one, and the following
section makes this debate moot.  However, we pedagogically consider
both cases in order to illustrate the issues that arise when one
treats the LW particles as an ideal gas.

We first calculate the partition function using the normalization
condition \eref{eq:rho_norm_1}.  It is convenient to perform the
standard transformations and work in a different basis (see, {\it
  e.g.},~\cite{Kapusta:1989}): One discretizes the momentum by
imposing periodic boundary conditions, and writes the Hamiltonian
$\hat{H} = \sum_{\bf p} \hat{h}_{\bf p}$ as a sum over the
single-particle Hamiltonians $\hat{h}_{\bf p} = E_{\bf p} \hat{N}_{\bf
  p}$.  The number operator $\hat{N}_{\bf p}$ has a spectrum
\begin{align}
	\hat{N}_{\bf p} \ket{n_{\bf q}} = n_{\bf p} \,
\delta_{{\bf p}, {\bf q}} \ket{n_{\bf p}} \, ,
\end{align}
where $\ket{n_{\bf p}}$ is the state containing $n_{\bf p}$ particles,
each of momentum ${\bf p}$.  In this basis, the partition function is
given by
\begin{align}\label{eq:Z_sloppy}
	Z = {\rm Tr} \, e^{- \beta \hat{H}} 
	= {\rm Tr} \, \prod_{\bf p} \, e^{- \beta E_{\bf p}
\hat{N}_{\bf p}} 
	= \prod_{\bf p} \sum_{n_{\bf p}=0}^{n_{\rm max}}
\expval{n_{\bf p}}{e^{- \beta E_{\bf p} \hat{N}_{\bf p}}}{n_{\bf p}}
\, , 
\end{align}
where $n_{\rm max} = \infty$ (1) for bosons (fermions).  Noting that
the norms are $\amp{n_{\bf p}}{n_{\bf p}} = (\eta_N)^{n_{\bf p}}$, one
finds
\begin{align}
	Z = \prod_{\bf p} \sum_{n_{\bf p}=0}^{n_{\rm max}}
\left( \eta_N e^{-\beta E_{\bf p}} \right)^{n_{\bf p}} \, .
\end{align}
Taking the logarithm turns the product into a sum, which becomes an
integral in the continuum limit.  Dividing by the volume factor, one
obtains the free energy density
\begin{align}\label{eq:thermo_freeenergydensity}
	\mathcal{F} = - (\beta V)^{-1} \ln Z = - \beta^{-1}
\int \frac{d^3p}{(2\pi)^3} \ln \left[ \sum_{n_{p}=0}^{n_{\rm max}}
\left( \eta_{N} e^{- \beta  E_{\bf p} } \right)^{n_{\bf p}} \right]
 \, .
\end{align}
The sum evaluated separately for bosons and fermions gives
\begin{align}
	\mathcal{F} = \begin{cases}
	 \beta^{-1} \int \frac{d^3p}{(2\pi)^3} \ln \left( 1 - \eta_{N}
\, e^{- \beta  E_{\bf p} } \right) & {\rm bosons} \, , \\
	 - \beta^{-1} \int \frac{d^3p}{(2\pi)^3} \ln \left( 1 +
\eta_{N} \, e^{- \beta  E_{\bf p} } \right) & {\rm fermions} \, ,
	 \end{cases}
\end{align}
which can be combined as
\begin{align}\label{eq:FreeEnergy_1}
	\mathcal{F} = \eta_{S} \beta^{-1} \int \frac{d^3p}{(2\pi)^3}
        \ln \left( 1 - \eta_{S} \eta_N e^{- \beta E_{\bf p}}  \right)
        \, ,
\end{align}
where $\eta_S = +1$ for bosons and $\eta_S = -1$ for fermions.  Had we
imposed the alternative normalization condition \eref{eq:rho_norm_2},
then the factor of $(\eta_N)^{n_{\bf p}}$ would not have arisen:
\begin{align}\label{eq:FreeEnergy_2}
	\mathcal{F}^{\prime} = \eta_{S} \beta^{-1} \int
\frac{d^3p}{(2\pi)^3} \ln \left( 1 - \eta_{S} e^{- \beta E_{\bf p}}
\right) \, ,
\end{align}
which is the standard free energy of an ideal gas.  

\begin{table}[t]
\begin{center}
\begin{tabular}{|l|c|c|c|c|c|}
\hline
 & $\eta_S$ & $\eta_N$ & $\beta^4 \mathcal{F}$ & $\beta^3 s$ &
$\beta^4 \rho$ \\ \hline
SM Boson & $+1$ & $+1$ & $+c_{0b} + c_{1b} \varepsilon$ &
$-4c_{0b} -2c_{1b} \varepsilon$ & $-3c_{0b} -c_{1b} \varepsilon$ \\
LW Boson (${\rm Tr}^{\prime} \hat{\rho} = 1$) & $+1$ & $-1$ &
$+c_{0b} +c_{1b} \varepsilon$ & $-4c_{0b} -2c_{1b} \varepsilon$ &
$-3c_{0b} -c_{1b} \varepsilon$ \\
LW Boson (${\rm Tr} \hat{\rho} = 1$)& $+1$ & $-1$ &
$-c_{0f} -c_{1f} \varepsilon$ & $+4c_{0f} +2c_{1f} \varepsilon$ &
$+3c_{0f} +c_{1f} \varepsilon$  \\
\hline
SM Fermion & $-1$ & $+1$ & $+c_{0f} +c_{1f} \varepsilon$ &
$-4c_{0f} -2c_{1f} \varepsilon$ & $-3c_{0f} -c_{1f} \varepsilon$ \\
LW Fermion (${\rm Tr}^{\prime} \hat{\rho} = 1$) & $-1$ & $-1$ &
$+c_{0f} +c_{1f} \varepsilon$ & $-4c_{0f} -2c_{1f} \varepsilon$ &
$-3c_{0f} -c_{1f} \varepsilon$ \\
LW Fermion (${\rm Tr} \hat{\rho} = 1$) & $-1$ & $-1$ &
$-c_{0b} -c_{1b} \varepsilon$ & $+4c_{0b} +2c_{1b} \varepsilon$ &
$+3c_{0b} +c_{1b} \varepsilon$  \\
\hline
\end{tabular}
\end{center}
\caption{
\label{tab:thermo_cases}
The thermodynamic properties of an ideal gas of SM or LW bosons or
fermions in the high-temperature limit $\beta^2 m^2 \equiv \varepsilon
\ll 1$.  For the LW particles, the density matrix is normalized using
either \eref{eq:rho_norm_1} or \eref{eq:rho_norm_2}, as indicated.
Higher-order terms in $\varepsilon$ are dropped.  The coefficients are
$c_{0b} \equiv -{\rm Li}_4 (+1)/\pi^2 = -\pi^2 /90$, $c_{1b} \equiv
{\rm Li}_2 (+1)/4\pi^2 \equiv 1/24$, $c_{0f} \equiv {\rm Li}_4
(-1)/\pi^2 = -7\pi^2 /720 = (7/8)c_{0b}$, and $c_{1f}
\equiv -{\rm Li}_2 (-1)/4\pi^2 = 1/48 = (1/2)c_{1b}$.}
\end{table}%

These results are summarized in \tref{tab:thermo_cases}, where we also
exhibit the entropy density $s = - \partial \mathcal{F} / \partial T$,
and energy density $\rho = \mathcal{F} + T s$.  We have expanded in
the high-temperature regime $\beta^2 m^2 \ll 1$ in order to facilitate
comparison with more familiar expressions.  When the density matrix is
normalized by summing the spectrum, ${\rm Tr}^{\prime} \hat{\rho} =
1$, one finds that the thermodynamics of a LW ideal gas is identical
to that of a SM ideal gas of the same spin.  This result is not
surprising, since the negative-norm property never enters.  On the
other hand, when the density matrix is normalized by taking
expectation values, ${\rm Tr} \, \hat{\rho} = 1$, one finds that the
LW boson has the same thermodynamics as a SM fermion with an overall
sign flip, and vice versa.  The negative entropy and energy densities
are a distinctly counterintuitive result, since the LW 1-particle
states have positive energy, and presumably should constitute an ideal
gas with positive energy density.  These results may be summarized
schematically as
\begin{align}
  {\rm Tr}^{\prime} \, \hat{\rho} = 1: &\qquad & \mathcal{F}
[\text{\small LW boson / fermion of mass $m$ }] =
+ \mathcal{F}[\text{\small SM boson / fermion of mass $m$ }]
\label{eq:LWgas_schematic_1} \, , \\
  {\rm Tr} \, \hat{\rho} = 1: &\qquad&\mathcal{F}
[\text{\small LW boson / fermion of mass $m$ }] =
- \mathcal{F}[\text{\small SM fermion /
    boson of mass $m$ }] \, .
\label{eq:LWgas_schematic_2}
\end{align}
Both of these results differ from a previous calculation of the
thermodynamics of LW ideal gas \cite{Bhattacharya:2011bb}, which finds
that the free energy, entropy, and energy densities of the LW gas are
precisely the opposite of these quantities for the corresponding SM
gas of the same spin, {\it i.e.},
\begin{align}
\begin{array}{l}
	\mathcal{F}[\text{\small LW boson / fermion of mass $m$ }] =
- \mathcal{F}[\text{\small SM boson / fermion of mass $m$ }]  
\end{array} \per
\end{align}
As noted above, this result agrees with that of Ref. \cite{Fornal:2009xc} derived in the LW resonance picture.  However, in Ref. \cite{Bhattacharya:2011bb} the authors assume that the positive-energy,
negative-norm LW particle states can be treated equivalently as states
of negative energy and positive norm.  It is not clear to us how a
thermodynamic system can have a spectrum of interacting particles 
which is unbounded both above (positive-energy states) 
and below (negative-energy states), nor can we justify the analytic continuation that is
required to define the partition function.  
As stated at the beginning of this section, the equivalent LW ideal gas approach with positive-energy, negative-norm states must also lead to instabilities; in this case, they arise through states of opposite norm combining to form zero-norm runaway modes \cite{Lee:1970iw, Boulware:1983vw}.  
Ultimately, we believe
that the formulation of LW theories, which forbids LW particles from
appearing as asymptotic states, is inconsistent with the picture that
LW particles form an ideal gas.

\subsection{LW Particles as Resonances}\label{sub:LWparticlesasResonanaces}

We now turn to the second picture of LW theories at finite
temperature, in which LW particles are not treated as fundamental
constituents of the gas.  By this we mean that, in the calculation of
expectation values $\langle \hat{\mathcal{O}} \rangle = {\rm Tr} \,
(\hat{\mathcal{O}} \hat{\rho})$, the trace extends over only the
subset of the Hilbert space containing states in which no LW particles
are present ({\it i.e.}, states annihilated by the LW particle
annihilation operators).  Instead, the LW fields make their presence
known through their interactions with the SM particles by modifying
the spectrum of the SM multi-particle states.  Treating these
interactions perturbatively, one can write the free energy density of a
LW theory schematically as
\begin{align}\label{eq:F_LWtheory}
	\mathcal{F}[\text{\small LW theory}] =
\mathcal{F}[\text{\small SM ideal gas}]  + \Delta \mathcal{F} \com
\end{align}
where the first term on the right-hand side represents the free energy
density of a ideal gas of SM particles, and the second term represents
perturbative corrections due to interactions among the SM and LW
fields.  Since the SM and LW fields interact through the SM gauge and
Yukawa couplings, the terms in $\Delta \mathcal{F}$ are the same order
as the so-called ``two-loop'' corrections in thermal field theory.
Generically, these corrections can be dropped in a leading-order
analysis.  However, as seen below, when the SM particles are able to
scatter through narrow-resonance LW particles, the corrections must be
resummed and become $O(1)$.

Before proceeding, note that one cannot apply the standard thermal
field-theory diagrammatic techniques to obtain $\Delta
\mathcal{F}$ (see, {\it e.g.}, \cite{Kapusta:1989}).  In this
formalism, one calculates the partition function by summing connected
graphs with no external lines using modified Feynman rules, so that
all of the fields are put on the same footing.  Here, however, one
needs to distinguish the SM particles, which can appear as external
states, and the LW particles, which are restricted to internal lines.

Fornal, Grinstein, and Wise~\cite{Fornal:2009xc} (FGW) studied a
scalar LW toy model at finite temperature, and we review their
calculation of the free energy density here.  In order to calculate
$\Delta \mathcal{F}$, FGW employed the formalism developed by Dashen,
Ma, and Bernstein~\cite{Dashen:1969ep} (DMB), by which the partition
function may be calculated from S-matrix elements.  DMB derived the
relationship
\begin{align}\label{eq:DeltaF_from_DMB}
	\Delta \mathcal{F} = - (\beta V)^{-1} \int dE \,
e^{- \beta E} \, \frac{1}{4\pi i} \left[ {\rm Tr} \, A S^{-1}(E)
\overleftrightarrow{ \frac{\partial}{\partial E} } S(E) \right]_c \, ,
\end{align}
where $S(E)$ is the S-matrix element between two multi-particle states
of energy $E$, $A$ symmetrizes (anti-symmetrizes) for bosons
(fermions), and $c$ denotes that only connected graphs are summed.  As
an example, FGW consider the scalar LW theory specified by the
Lagrangian
\begin{align}\label{eq:Ltoy_FGW}
	\mathcal{L} = \frac{1}{2} (\partial_{\mu} \hat{\phi})^2 -
\frac{1}{2M^2} (\partial^2 \hat{\phi})^2 - \frac{1}{2} m^2
\hat{\phi}^2 - \frac{g}{3!} \hat{\phi}^3 \per
\end{align}
The interaction term $g \hat{\phi}^3$ allows a LW particle to decay
into two SM particles with a width given by
\begin{align}
	\Gamma = \frac{-g^2}{32 \pi M} \sqrt{1 - \frac{4 m^2}{M^2} }
\per
\end{align}
The width is negative because of the negative residue of the LW
propagator~\cite{Grinstein:2007mp}.  The same interaction allows two
SM particles to scatter through a LW particle resonance, with matrix
element
\begin{align}\label{eq:propagator_FGW}
	\mathcal{M} = \frac{1}{2} \cdot
\frac{-g^2}{E^2 - {\bf P}^2 - M^2 + i M \Gamma} \, ,
\end{align}
where $S(E) = 1 - i \mathcal{T}(E)$ and 
\begin{align}
	\expval{{\bf p}_1 \, {\bf p}_2}{\mathcal{T}(E)}{{\bf q}_1 \,
{\bf q}_2} = (2\pi) \delta (E - E_1 - E_2) (2\pi)^3 \delta({\bf p}_1
+ {\bf p_2} - {\bf q}_1 - {\bf q}_2) \mathcal{M}(E) \per
\end{align}
Upon evaluating \eref{eq:DeltaF_from_DMB} and taking the narrow-width
approximation $\Gamma \ll M$, FGW find
\begin{align}\label{eq:DeltaF_FGW}
	\Delta \mathcal{F} = - \beta^{-1} \int \frac{d^3 p}{(2\pi)^3}
\, \ln \left( 1 - e^{- \beta \sqrt{{\bf p}^2 + M^2}} \right) \per
\end{align}
This is precisely the form of the free energy density of an ideal gas
of bosons, but with an overall minus sign [{\it cf.}
\eref{eq:FreeEnergy_1} with $\eta_S = \eta_N = +1$].  At least, the
fact that $\Delta \mathcal{F}$ takes the form of an ideal gas term is
reassuring; in the narrow-width approximation the resonances are
long-lived, and contribute to the free energy as if they were stable
constituents of the plasma~\cite{Dashen:1969ep, Dashen:1974jw}.  On
the other hand, the minus sign is surprising.  We have already seen
that the calculation of the free energy density of an ideal gas of LW
bosons produces one of the two results in \tref{tab:thermo_cases}, and
neither of these correspond to \eref{eq:DeltaF_FGW}.  The minus sign
appears because $\Gamma < 0$, and the limit $\Gamma \to 0^-$ of
Eq.~(\ref{eq:propagator_FGW}) at its pole differs from the limit
$\Gamma \to 0^+$ by a sign.  In other words, the free energy density
is nonanalytic at $\Gamma \propto g^2 = 0$.  FGW generalize their
result from the scalar toy model to also consider the fermionic LW
resonances, and they find the same overall sign flip.  We summarize
this result by setting a sign placeholder $\sigma = -1$ in
\begin{align}\label{eq:LWsignflip}
	\left. \Delta \mathcal{F}[\text{\small LW boson/fermion
	narrow resonance of mass M}] = \sigma \mathcal{F}[\text{\small
	SM boson/fermion ideal gas of mass M}] \right|_{\sigma = -1}
\, ,
\end{align}
to which we refer as the ``LW sign flip.''

One may worry that the S-matrix formalism given by
\eref{eq:DeltaF_from_DMB} is inapplicable to the study of LW theories,
for instance because the negative-norm states were not properly taken
into account in the derivation of DMB or FGW\@.  After a careful
review of the calculations in those works, we can find no obvious
source of error.  Nevertheless, it was pointed out by Espinosa and
Grinstein~\cite{Espinosa:2011js} (EG) that the result
\eref{eq:LWsignflip} leads to unexpected breakdown of the well-known
connection between UV behavior and symmetry restoration.  This
connection derives from the fact that the graphs giving rise to
quadratic divergences at $T=0$ are the same graphs responsible for
$O(T^2)$ self-energy corrections at finite temperature
\cite{Comelli:1996vm}.  For example, if a bosonic field has a
divergent self-energy correction $\Delta m^2 =
\kappa \Lambda^2 / 16 \pi^2$, then it receives a thermal mass
correction $\Delta m^2 = \kappa T^2 / 12$, and for fermions one has
$\Delta m^2 = - \kappa \Lambda^2 / 16 \pi^2$ and $\Delta m^2 = -\kappa
T^2 / 24$.  In models that solve the hierarchy problem by a
cancellation of quadratic divergences between degrees of freedom of
the same spin, this connection implies that there should also be a
cancellation of the leading thermal mass corrections.  However, if the
$\sigma = -1$ LW sign flip in \eref{eq:LWsignflip} is the correct result,
then there is no such cancellation (see \sref{sub:Veff_LW_Toy}).
Instead, to obtain the cancellation, the sign of the LW correction
must be the same as that of the corresponding SM partner, the $\sigma
= +1$ case of
\begin{align}\label{eq:LWsamesign}
	\left. \Delta \mathcal{F}[\text{\small LW boson/fermion
	narrow resonance of mass M}] = \sigma \mathcal{F}[\text{\small
	SM boson/fermion ideal gas of mass M}] \right|_{\sigma = +1}
	\per
\end{align}
In models with spontaneously broken symmetries, this effect tends to
retard symmetry restoration.

We present a simple, heuristic argument based on energetics that lends
credence to the result \eref{eq:DeltaF_FGW}.  Recall that the free
energy density is given by $\mathcal{F} = - (\beta V)^{-1} \ln {\rm
Tr} \, e^{- \beta \hat{H}}$, where the trace extends over the states
containing multiple SM particles and no LW particles.  In the absence
of interactions, the SM particles are free, and one obtains the ideal
gas term in \eref{eq:F_LWtheory}.  The interactions affect
$\mathcal{F}$ by changing the energy of the multi-particle states.
For example, consider the theory \eref{eq:Ltoy_FGW} studied by FGW\@.
Two SM scalars may interact by the exchange of a SM virtual particle.
The scalar field mediates an attractive force characterized by the
Yukawa potential, which lowers the energy of the two-particle state
and yields $\Delta \mathcal{F} < 0$.  On the other hand, when the two
SM scalars exchange a LW virtual particle, the propagator
\eref{eq:propagator_FGW} gives rise to a repulsive force which, in
turn, raises the energy of the two-particle state and yields $\Delta
\mathcal{F} > 0$ as in \eref{eq:DeltaF_FGW} (since the logarithm
is always negative).  This argument does not confirm the form of
\eref{eq:DeltaF_FGW}, but it does suggest that the sign flip in
Eq.~(\ref{eq:LWsignflip}) may be correct.

Let us now summarize.  The question of which picture provides the
correct description of LW theories at finite temperature remains
unsettled.  We have argued that treating the LW particles as
resonances appears to be more consistent with the boundary conditions
that protect LW theories from the pathologies that generally plague HD
theories.  However, some uncertainty remains as to the sign of $\Delta
\mathcal{F}$, as contrasted in Eqs.~(\ref{eq:LWsignflip}) and
(\ref{eq:LWsamesign}).  In order to keep our analysis as general as
possible, we consider both possibilities by maintaining the index
$\sigma = \pm1$ as a prefactor to LW field contribution to the
effective potential, and study both cases simultaneously.  Despite
this effort to remain completely agnostic, it should be noted that an
entirely different third possibility is not excluded.

\subsection{Thermal Effective Potential of a LW Toy Model}
\label{sub:Veff_LW_Toy}

Before proceeding to evaluate the thermal effective potential for the
LWSM, we begin by considering a pair of LW toy theories.  We do not
reproduce here the derivation of the one-loop thermal effective
potential since, apart from the modifications to the thermal
correction discussed above, the derivation is standard (for a review
see \cite{Quiros:1999jp}).  To wit, one extends the Lagrangian by introducing a source term 
(tadpole) for the scalar field, calculates the partition function in the 
presence of this source, and performs a Legendre transformation to 
express the source in terms of the expectation value of the scalar field, 
$\phi_c$.  The thermal effective potential is simply the $\phi_c$--dependent 
free energy density that can be obtained from the logarithm 
of the partition function.  This calculation may be performed using a 
number of techniques, such as a diagrammatic approach and the 
path integral formalism, and each admits a perturbative expansion.  At ``one-loop'' order, the effective
potential may be written as the sum
\begin{align}\label{eq:Veff_U_0_T}
	V_{\rm eff}^{\rm (1L)}(\phi_c) = U(\phi_c) + \Delta V_{0}^{\rm
(1L)}(\phi_c) + \Delta V_{T}^{\rm (1L)}(\phi_c, T) \, ,
\end{align}
where $\phi_c$ is the scalar condensate and $T$ the temperature.
These three terms correspond, respectively, to the classical potential
energy $U(\phi_c)$, the non-thermal correction
\begin{align}\label{eq:DeltaV0}
	\Delta V_{0}^{\rm (1L)}(\phi_c) = \delta V_{\rm c.t.}(\phi_c)
+ \begin{cases}
	\frac{1}{2} \sum_b \int \frac{d^4 p_E}{(2\pi)^4}
\ln \left[ p_E^2 + {m}_b^2(\phi_c) \right]  & {\rm bosons} \, ,
\\
	- \sum_f \int \frac{d^4 p_E}{(2\pi)^4}
\ln \left[ p_E^2 + {m}_f^2(\phi_c) \right] & {\rm fermions} \, ,
	\end{cases}
\end{align}
arising from renormalized quantum vacuum fluctuations of the various
fields in the theory (the counterterms are contained in $\delta
V_{\rm c.t.}$), and the thermal correction
\begin{align}\label{eq:DeltaVT}
	\Delta V_{T}^{\rm (1L)}(\phi_c,T) = \begin{cases}
	\sum_b \, T \int \frac{d^3p}{(2\pi)^3}
\ln \left( 1 - e^{-\beta \sqrt{{\bf p}^2 + {m}_b^2(\phi_c)}}
\right) & {\rm bosons} \, , \\
	-\sum_f \, T \int \frac{d^3p}{(2\pi)^3} \ln \left( 1 +
e^{-\beta \sqrt{{\bf p}^2 + {m}_f^2(\phi_c)}} \right) &
{\rm fermions} \, ,
	\end{cases}
\end{align}
arising from the presence of a gas of free particles, in accord with
Eq.~(\ref{eq:FreeEnergy_2}).  The sums run over the various bosonic
($b$) and fermionic ($f$) fields in the theory under consideration.
The functions ${m}_{b,f}^2(\phi_c)$ represent the effective masses of
the various fields of the theory in the presence of the condensate
$\phi_c$.  At one-loop order, the effective potential only depends
upon these masses and $U(\phi_c)$.  One need make only minimal
modifications to these expressions to account for the negative-norm LW
fields.  As discussed above, the leading-order thermal corrections
arising from the LW fields are of the same form as the ideal gas term
\eref{eq:DeltaVT}, and only the overall sign is under dispute.  We
remind the reader of the index $\sigma = \pm 1$ used to consider both
cases [Eqs.~(\ref{eq:LWsignflip})--(\ref{eq:LWsamesign})]
simultaneously.  The non-thermal corrections to the one-loop effective
potential due to LW fields were calculated previously by
\cite{Espinosa:2011js}.  Unsurprisingly, the result is identical to
\eref{eq:DeltaV0}.  Since the quantum effective potential is merely a
sum of zero-point energies $\sum_{\bf p} \frac{1}{2} \hbar \omega_{\bf
  p}$, there is no relative sign difference between the SM and LW
field contributions to $\Delta V_0^{\rm (1L)}$.  We now use
\eref{eq:Veff_U_0_T} with the $\sigma$-factor modifications to
evaluate the effective potential for two LW toy theories.

\subsubsection{A Scalar Example}
First, consider the toy theory of an interacting, real scalar field
$\hat{\phi}(x)$ described by the Lagrangian
\begin{align}\label{eq:LWscalartoy_Lagrangian}
	\mathcal{L} &= - \frac{1}{2\LamLW^2}
( \partial^2 \hat{\phi} )^2 + \frac{1}{2}
(\partial_{\mu} \hat{\phi})^2 - U(\hat{\phi}) \, , \nn
	U(\hat{\phi}) &= \Omega + \frac{1}{2} \mu^2 \hat{\phi}^2 +
\frac{\lambda}{4} \hat{\phi}^4 \per
\end{align}
Let $\phi_c = \langle \hat{\phi} \rangle$ be the homogeneous
condensate, and expand $\hat{\phi}(x) = \phi_c + \hat{\varphi}(x)$.
To obtain the effective mass of the field $\hat{\varphi}$, we expand
the Lagrangian to quadratic order and find
\begin{align}\label{eq:LWscalartoy_Lagrangian_2}
	\mathcal{L} \supset 
	- \frac{1}{2} \hat{\varphi} \left( \frac{\partial^4}{\LamLW^2}
+ \partial^2 + {m}_{\hat{\varphi}}^2(\phi_c) \right)
\hat{\varphi} \, ,
\end{align}
where
\begin{align}
	{m}_{\hat{\varphi}}^2(\phi_c) \equiv U^{\prime \prime}
(\phi_c) = \mu^2 + 3 \lambda \phi_c^2 \per
\end{align}
As discussed in \sref{sec:Intro_to_LW}, the higher-order derivative in
\eref{eq:LWscalartoy_Lagrangian_2} implies that the field
$\hat{\varphi}$ carries two degrees of freedom.  To disentangle them,
and thereby identify the SM and LW component fields, one may Fourier
transform to obtain the propagator,
\begin{align}\label{eq:LWscalartoy_propagator}
	D_{\hat{\varphi}}(p) = i \left( - \frac{p^4}{\LamLW^2} + p^2 -
{m}_{\varphi}^2(\phi_c) \right)^{-1} =
\frac{\LamLW^2}{{m}_{\tilde{\varphi}}^2 - {m}_{\varphi}^2} \left(
\frac{i}{p^2 - {m}_{\varphi}^2} - \frac{i}{p^2 -
{m}_{\tilde{\varphi}}^2} \right) \, ,
\end{align}
where
\begin{align}\label{eq:LWscalartoy_poles}
\begin{array}{lcl}
	\text{Positive-Norm Pole:} & \qquad & {m}_{\varphi}^2(\phi_c)
\equiv \frac{\LamLW^2}{2} \left( 1 - \sqrt{ 1 -
\frac{4 {m}_{\hat{\varphi}}^2(\phi_c)}{\LamLW^2} } \right) \, , \\
	\text{Negative-Norm Pole:} & \qquad &
{m}_{\tilde{\varphi}}^2(\phi_c) \equiv \frac{\LamLW^2}{2}
\left( 1 + \sqrt{ 1 -
\frac{4 {m}_{\hat{\varphi}}^2(\phi_c)}{\LamLW^2} } \right) \, ,
\end{array}
\end{align}
Essential to this decomposition is a consistent pole prescription and
direction of Wick rotation, as discussed in App.~\ref{sec:quant_conv}.
It is now straightforward to construct the one-loop effective
potential using \eref{eq:Veff_U_0_T}.  One finds
\begin{align}\label{eq:LWscalartoy_Veff}
	V^{\rm (1L)}_{\rm eff}(\phi_c,T) =\ & U(\phi_c) +
\left[ \delta V_{\rm c.t.}(\phi_c) + \frac{1}{2} \int
\frac{d^4 p_E}{(2\pi)^4} \Bigl( \ln [p_E^2 + {m}_{\varphi}^2(\phi_c) ]
+ \ln [p_E^2 + {m}_{\tilde{\varphi}}^2(\phi_c) ] \Bigr) \right] \nn
	& + \frac{T^4}{2 \pi^2} \Bigl[ J_{B}({m}_{\varphi}^2 / T^2) +
\sigma J_{B}({m}_{\tilde{\varphi}}^2 / T^2) \Bigr] \, ,
\end{align}
where we have defined the bosonic thermal function
\begin{align}\label{eq:JB_def}
	J_{B}(y) \equiv \int_{0}^{\infty} dx \, x^2 \, \ln
\left( 1 - e^{- \sqrt{x^2 + y} } \right) \com
\end{align}
and have introduced the index $\sigma = \pm 1$ in front of the
thermal correction arising from the LW field.  The counterterms
\begin{align}\label{eq:LWscalartoy_deltaVct}
	\delta V_{\rm c.t.}(\phi_c) = \delta \Omega + \frac{1}{2}
\delta \mu^2 \phi_c^2 + \frac{\delta \lambda}{4} \phi_c^4 \, ,
\end{align}
are determined once a set of renormalization conditions are specified.

\subsubsection{A Fermionic Example}
As a second example, consider the toy LW theory
\begin{align}
	\mathcal{L} = \bar{\hat{\psi}}
\left( i\frac{\slashed{\partial}^3}{\LamLW^2} + i \slashed{\partial} -
\lambda \hat{\phi} \right) \hat{\psi} + \mathcal{L}_{\hat{\phi}} \, ,
\label{eq:LWfermiontoy_L}
\end{align}
in which the Dirac spinor $\hat{\psi}$ acquires its mass through the
Yukawa coupling with the real scalar field $\hat{\phi}$.  The term
$\mathcal{L}_{\hat{\phi}}$ contains the kinetic term, mass term, and
interactions for $\hat{\phi}$.  The details of these interactions are
not relevant, and we merely suppose that they are such that a
condensate $\langle \hat{\phi} \rangle = \phi_c$ forms.  It is
straightforward to analyze the propagator for $\hat{\psi}$:
\begin{align}
	D_{\hat{\psi}}(p) = \ & i \left(
-\frac{\slashed{p}^3}{\LamLW^2} + \slashed{p} -
{m}_{\hat{\psi}}(\phi_c) \right)^{-1} \nn
	= \ & + \frac{\LamLW^2}{({m}_{\tilde{\psi}_1} - {m}_{\psi})
({m}_{\psi} - {m}_{\tilde{\psi}_2})} \cdot
\frac{i}{\slashed{p} - {m}_{\psi}} \nn
	&- \frac{\LamLW^2}{({m}_{\tilde{\psi}_1} - {m}_{\psi})
({m}_{\tilde{\psi}_1} - {m}_{\tilde{\psi}_2})} \cdot
\frac{i}{\slashed{p} - {m}_{\tilde{\psi}_1}}  \nn
	&- \frac{\LamLW^2}{({m}_{\psi} - {m}_{\tilde{\psi}_2})
({m}_{\tilde{\psi}_1} - {m}_{\tilde{\psi}_2})} \cdot
\frac{i}{\slashed{p} - {m}_{\tilde{\psi}_2}} \, , 
\end{align}
where
\begin{align}\label{eq:LWfermiontoy_poles}
\begin{array}{lcl}
	\text{Positive-Norm Pole:} & \qquad & {m}_{\psi}(\phi_c)
\equiv \LamLW \sqrt{\frac{2}{3} \left( 1 - \cos \frac{\theta}{3}
\right)} \, , \\
	\text{Negative-Norm Pole:} & \qquad &
{m}_{\tilde{\psi}_1}(\phi_c) \equiv \LamLW
\sqrt{ \frac{2}{3} \left( 1 + \cos \frac{\theta + \pi}{3} \right) } \, , \\
	\text{Negative Norm Pole:} & \qquad &
{m}_{\tilde{\psi}_2}(\phi_c) \equiv - \LamLW \sqrt{ \frac{2}{3}
\left( 1 + \cos \frac{\theta - \pi}{3} \right) } \, , 
\end{array} 
\end{align}
and
\begin{align}
\begin{array}{ll}
	\theta & \equiv \arctan \,
\frac{2 \sqrt{\alpha(1-\alpha)}}{1-2 \alpha} \, , \hspace{1em} 0 \le
\theta < \pi \, , \\
	\alpha & \equiv \frac{27}{4} 
\frac{{m}_{\hat{\psi}}^2}{\LamLW^2} \, , \\
	{m}_{\hat{\psi}} & \equiv \lambda \phi_c \, .
\end{array}
\end{align}
As before, one can construct the one-loop effective potential
using \eref{eq:Veff_U_0_T}:
\begin{align}\label{eq:LWfermiontoy_Veff}
	V^{\rm (1L)}_{\rm eff}(\phi_c,T) & = U(\phi_c) \nn & +
\left[ \delta V_{\rm c.t.}(\phi_c) - \int \frac{d^4 p_E}{(2\pi)^4}
\Bigl( \ln [p_E^2 + {m}_{\psi}^2(\phi_c) ] +
\ln [p_E^2 + {m}_{\tilde{\psi}_1}^2(\phi_c) ] +
\ln [p_E^2 + {m}_{\tilde{\psi}_2}^2(\phi_c) ] \Bigr) \right] \nn
	& - \frac{T^4}{2 \pi^2} \Bigl[ J_{F}({m}_{\psi}^2 / T^2) +
\sigma J_{F}({m}_{\tilde{\psi}_1}^2 / T^2) +
\sigma J_{F}({m}_{\tilde{\psi}_2}^2 / T^2) \Bigr] \, ,
\end{align}
where we have defined the fermionic thermal function 
\begin{align}\label{eq:JF_def}
	J_{F}(y) \equiv \int_{0}^{\infty} dx \, x^2 \,
\ln \left( 1 + e^{- \sqrt{x^2 + y} } \right) \per
\end{align}
Once again, the counterterms in $\delta V_{\rm c.t.}$ are determined
by renormalization.

\subsubsection{Comparison of Bosonic and Fermionic Cases}
Let us now pause to comment on the results \eref{eq:LWscalartoy_Veff}
and \eref{eq:LWfermiontoy_Veff}.  First, note that the scalar masses
\eref{eq:LWscalartoy_poles} are only real-valued for $\phi_c^2 <
(\LamLW^2 - 4 \mu^2) / 12 \lambda$, and the fermion masses
\eref{eq:LWfermiontoy_poles} only for $\phi_c^2 < 4 \LamLW^2 / 27
\lambda^2$.  In either case, if the value of $\phi_c$ becomes too
large, then the LW stability condition discussed in
\sref{sec:Intro_to_LW} breaks down.  Since the thermal corrections
tend to give rise to symmetry restoration, then $\phi_c \to 0$,
and we do not concern ourselves with the failure of the calculation at
large $\phi_c$.  At small $\phi_c$ the masses can be expanded as
\begin{subequations}\label{eq:mass_expansions}
\begin{align}
	{m}_{\varphi}^2(\phi_c) &\approx {m}_{\hat{\varphi}}^2 +
\frac{{m}_{\hat{\varphi}}^4}{\LamLW^2} +
O({m}_{\hat{\varphi}}^6 / \LamLW^4)  \label{eq:expand_mphi} \, ,\\
	{m}_{\tilde{\varphi}}^2(\phi_c) &\approx \LamLW^2 -
{m}_{\hat{\varphi}}^2 - \frac{{m}_{\hat{\varphi}}^4}{\LamLW^2}
+ O({m}_{\hat{\varphi}}^6 / \LamLW^2) \label{eq:expand_mphiLW} \, ,
\\
	{m}_{\psi}^2(\phi_c) &\approx {m}_{\hat{\psi}}^2 +
2\frac{{m}_{\hat{\psi}}^4}{\LamLW^2} +
O({m}_{\hat{\psi}}^6 / \LamLW^4)  \, , \label{eq:expand_mpsi} \\
	{m}_{\tilde{\psi}_{1,2}}^2(\phi_c) & \approx \LamLW^2
\mp {m}_{\hat{\psi}} \LamLW - \frac{1}{2} {m}_{\hat{\psi}}^2
\mp \frac{5}{8} \frac{ {m}_{\hat{\psi}}^3 }{\LamLW} -
\frac{{m}_{\hat{\psi}}^4}{\LamLW^2} +
O({m}_{\hat{\psi}}^6 / \LamLW^4)  \per \label{eq:expand_mpsiLW}
\end{align}
\end{subequations}
For the LW field masses, Eqs.~(\ref{eq:expand_mphiLW}) and
(\ref{eq:expand_mpsiLW}), the $\phi_c$ dependence (carried by
${m}^2_{\hat{\varphi}}$ and ${m}^2_{\hat{\psi}}$) is subdominant to
the LW scale $\LamLW$.  Therefore, one expects that these fields just
give a constant ($\phi_c$-independent) shift to the effective
potential except at field values at which ${m}_{\hat{\varphi}}^2,
{m}_{\hat{\psi}}^2 \sim \LamLW^2$.  It is worth emphasizing that the
issue here is not simply that the LW fields are too heavy and
decouple, but rather that they acquire their mass through a constant
mass parameter instead of entirely through symmetry breaking.  In this
way, their impact on the effective potential is comparable to that of
heavy squarks in supersymmetric theories.

Since the LW stability condition requires $\LamLW$ to remain larger
than the $\phi_c$-dependent mass scales, one expects that the LW
fields provide negligible contributions to the non-thermal effective
potential $\Delta V_0^{\rm (1L)}$.  Similarly, at low temperatures
$T^2 \ll \LamLW^2$, the LW fields are heavy and their contributions to
the thermal effective potential are Boltzmann suppressed since $J_B(y)
\sim J_F(y) \sim e^{-\sqrt{y}}$ for $y \gg 1$.  For theories with spontaneous
symmetry breaking, such as the scalar toy model
\eref{eq:LWscalartoy_Lagrangian} with $\mu^2 < 0$, one expects the LW
fields to have a negligible impact on symmetry restoration and the
phase transition unless the phase transition temperature $T_c$ is
comparable to $\LamLW$.  Naturally, $T_c$ is set by the mass scale of
the field experiencing the phase transition, so that $T_c \sim
{m}_{\hat \varphi}(v)$, where $\phi_c = v$ is the zero-temperature
scalar vacuum expectation value.  However, due to the LW stability
condition $\LamLW > 2 {m}_{\hat \varphi}(v)$, one sees that the limit
$T_{c} \to \LamLW$ cannot be reached.  We reach the general conclusion
that, in natural scenarios, the LW fields do not qualitatively affect
the phase transition.

On the other hand, at high temperatures $T^2 \gtrsim \LamLW^2$, the
thermal corrections $\Delta V_T^{\rm (1L)}$ can become significant.
In the high-temperature limit, the bosonic and fermionic thermal
functions admit the series expansions~\cite{Kapusta:1989}
\begin{align}\label{eq:JB_highT}
	&J_B(y) \xrightarrow{y \ll 1} - \frac{\pi^4}{45} +
\frac{\pi^2}{12} y - \frac{\pi}{6} y^{3/2} - \frac{1}{32} y^2 \ln
\frac{y}{a_b} + O(y^3) \, , \\ \label{eq:JF_highT}
	&J_F(y) \xrightarrow{y \ll 1} + \frac{7\pi^4}{360} -
\frac{\pi^2}{24} y - \frac{1}{32} y^2 \ln \frac{y}{a_f} + O(y^3) \, ,
\end{align}
where $a_b = 16 a_f = 16 \pi^2 \exp{3/2 - 2 \gamma_E}$.  The thermal
correction in Eqs.~(\ref{eq:LWscalartoy_Veff}),
(\ref{eq:LWfermiontoy_Veff}) can now be expanded using
Eqs.~(\ref{eq:JB_highT})--(\ref{eq:JF_highT}).  First, for the scalar
toy theory, one finds:
\begin{align}\label{toybosonenergy}
	\frac{T^4}{2 \pi^2} \left[ J_B \left( \frac{m_{\varphi}^2(\phi_c)}{T^2} \right) + \sigma J_B \left( \frac{m_{\tilde{\varphi}}^2(\phi_c)}{T^2}  \right) \right]
	\xrightarrow{T^2 \gg \LamLW^2 \gg m_{\hat{\varphi}}^2} - \frac{\pi^2}{90} (1 + \sigma) T^4 + \frac{\LamLW^2 T^2}{24} \sigma + (1-\sigma) \frac{{m}_{\hat{\varphi}}^2 T^2}{24} \nn
	\hspace{2.2cm} - \sigma \frac{\LamLW^3 T}{12 \pi} + \sigma \frac{\LamLW T {m}_{\hat{\varphi}}^2}{8 \pi} - \frac{T}{12 \pi} ({m}_{\hat{\varphi}}^2)^{3/2} + \ldots \per
\end{align}
The first term is the free energy density of a relativistic gas with
$(1+\sigma)$ degrees of freedom.  If one takes $\sigma = -1$, then
this term vanishes due to a cancellation between the two degrees of
freedom.  In this case, the leading temperature dependence is given by
the $O(T^2)$ term, and consequently the thermodynamic quantities, such
as the equation of state, are modified from the familiar expressions
for radiation domination \cite{Fornal:2009xc}.  The third term carries
the field dependence and is responsible for symmetry restoration.
Since $m_{\hat \varphi}^2 \sim \phi_c^2$, this term is an effective
temperature-dependent mass for the scalar $\varphi$.  For the case
$\sigma = +1$, there is a cancellation between the SM and LW fields,
and this term vanishes.  Then symmetry restoration is accomplished by
the $O(\LamLW T)$ term [relative size $O(T/\LamLW)$], which tends to
increase the temperature of symmetry restoration.  For the fermionic
toy theory, on the other hand, one finds:
\begin{align}\label{toyfermionenergy}
	-\frac{T^4}{2 \pi^2} & \left[ J_F \left( \frac{m_{\psi}^2(\phi_c)}{T^2} \right) + \sigma J_F \left( \frac{m_{\tilde{\psi}_1}^2(\phi_c)}{T^2} \right) + \sigma J_F \left( \frac{m_{\tilde{\psi}_2}^2(\phi_c)}{T^2} \right) \right] \nn
	\xrightarrow{T^2 \gg \LamLW^2 \gg m_{\hat{\psi}}^2} & - \frac{7\pi^2}{720} (1 + 2\sigma) T^4 + 2\sigma \frac{\LamLW^2 T^2}{48} + (1-\sigma) \frac{m_{\hat{\psi}}^2 T^2}{48} + (1 - \sigma) \frac{ m_{\hat{\psi}}^4 T^2}{24 \LamLW^2}  \nn
	& + \frac{{m}_{\hat{\psi}}^4}{64 \pi^2} \left( \ln \frac{{m}_{\hat{\psi}}^2}{a_f T^2} - \sigma \ln \frac{ \LamLW^2}{a_f T^2} \right) + 2 \sigma \frac{\LamLW^4}{64 \pi^2} \ln \frac{\LamLW^2}{a_f T^2} 
	+ \ldots \per
\end{align}
In this case, the leading $O(T^4)$ term flips sign for $\sigma = -1$.
Since the free energy density and pressure carry opposite signs, the
system develops a negative pressure for $T \gg \LamLW$.  Once again,
the $O(m^2 T^2)$ term vanishes in the case $\sigma = +1$, but now the
next [$O(m_{\hat{\psi}}^4 T^2)$] term that could restore symmetry
vanishes as well, since the nonanalytic $y^{3/2}$ term of $J_B$ is
absent from $J_F$.  Symmetry restoration must be accomplished in the
bosonic sector.

\section{The LWSM at Finite Temperature}\label{sec:LWSM_finiteT}

\subsection{The LWSM Thermal Effective Potential}\label{sub:LWSM_Veff}

In this section we construct the thermal effective potential for the
LWSM using the results of \sref{sec:Thermo_of_LW}.  As seen there, to
calculate the effective potential in the one-loop approximation, one
sums the separate contributions arising from each of the SM fields and
its LW partner.  The fields that couple more strongly to the Higgs
give a larger contribution to $V_{\rm eff}$.  Thus, to a very good
approximation, one need only sum the contributions arising from the
top quark, the weak gauge bosons, the Higgs boson, and each of their
LW partners.  To ensure a correct counting of the degrees of freedom,
we assume that the remaining SM-like degrees of freedom are massless,
that the remaining LW-like degrees of freedom have mass $\LamLW^2$,
and we include three ``SM gauge ghosts'' that contribute negative
degrees of freedom, as per the discussion in
App.~\ref{sec:Derive_Masses}.

We parametrize the Higgs condensate as $\langle \hat{H} \rangle = ( 0
\, , \, \phi_c / \sqrt{2} )^T$, where $\phi_c = v = 246 \GeV$
corresponds to the tree-level vacuum expectation value, calculate the field-dependent
masses in \aref{sec:Derive_Masses}, and summarize the results in
\tref{tab:LWSM_masses}.  For each species, we list the spin $s$, the
number of dynamical degrees of freedom $g$ (arising from color, spin,
and isospin), and the field-dependent squared mass $m^2(\phi_c)$.  In
light of the discussion of the preceding section, we allow for
ambiguity in the sign of the thermal corrections to the effective
potential by introducing an index $\sigma$ that equals $1$ for the
SM-like fields and either $\pm 1$ for the LW-like fields.  Finally, as
discussed in \sref{sub:Veff_LW_Toy}, the field-dependent mass
eigenvalues become complex for large values of $\phi_c$ at which the
LW stability condition fails, and the system becomes unstable; since
we are primarily interested in the regime $\phi_c < v$, we ignore such
cases.

\begin{table}[t]
\begin{center}
\begin{tabular}{|c|c|c|c|c|c|c|}
\hline
Field & $s$ & $g$ &  $\sigma$ & \multicolumn{2}{|c|}{${m}_i^2(\phi_c)$}  \\ \hline
SM-like Higgs & 0 & 1 & 1 & 
	${m}_h^2 = \frac{1}{2} \LamH^2 \left( 1 - \sqrt{1 - \frac{4 {m}^2_{\hat{h}}}{\LamH^2}} \right)$ & 
	\multirow{2}{*}{${m}_{\hat{h}}^2 = \lambda ( 3 \phi_c^2 - v^2 )$} \\
LW-like Higgs & 0 & 1 & $\sigma$ & 
	${m}_{\tilde{h}}^2 = \frac{1}{2} \LamH^2 \left( 1 + \sqrt{1 - \frac{4 {m}^2_{\hat{h}}}{\LamH^2}} \right)$ & 
	\\
SM-like pseudoscalar & 0 & 1 & 1 & 
	${m}_{P}^2 = \frac{1}{2} \LamH^2 \left( 1 - \sqrt{1 - \frac{4 {m}^2_{\hat{P}}}{\LamH^2}} \right) $ & 
	\multirow{2}{*}{${m}_{\hat{P}}^2 = \lambda ( \phi_c^2 - v^2 )$} \\ 
LW-like pseudoscalar & 0 & 1 & $\sigma$ & 
	${m}_{\tilde{P}}^2 = \frac{1}{2} \LamH^2 \left( 1 + \sqrt{1 - \frac{4 {m}^2_{\hat{P}}}{\LamH^2}} \right) $ & 
	\\ 
SM-like charged scalar & 0 & 2 & 1 & 
	${m}^2_{h^{\pm}} = \frac{1}{2} \LamH^2 \left( 1 - \sqrt{1 - \frac{4 {m}^2_{\hat{h}^{\pm}}}{\LamH^2}} \right) $ 
	& \multirow{2}{*}{${m}_{\hat{h}^{\pm}}^2 = \lambda ( \phi_c^2 - v^2 )$} \\ 
LW-like charged scalar & 0 & 2 & $\sigma$ & 
	${m}^2_{\tilde{h}^{\pm}} = \frac{1}{2} \LamH^2 \left( 1 + \sqrt{1 - \frac{4 {m}^2_{\hat{h}^{\pm}}}{\LamH^2}} \right) $ & 
	\\ 
SM gauge ghosts & 0 & $-3$ & 1 & 0 & 
	\\ \hline
SM-like $W$ & 1 & 6 & 1 & 
	${m}^2_{\tilde{W}^{\pm}} \! = \frac{1}{2} \LamW^2 \left( 1 - \sqrt{1 - \frac{4 {m}^2_{\hat{W}}}{\LamW^2}} \right)$ & 
	\multirow{2}{*}{${m}_{\hat{W}^{\pm}}^2 = \frac{g^2 \phi_c^2}{4} $} \\
LW-like $W$ & 1 & 6 & $\sigma$ & 
	${m}^2_{\tilde{W}^{\pm}} \! = \frac{1}{2} \LamW^2 \left( 1 + \sqrt{1 - \frac{4 {m}^2_{\hat{W}}}{\LamW^2}} \right)$ & 
	\\ \hline
SM-like $A$ & 1 & 2 & 1 & 
	${m}^2_{A} = 0$ & 
	\\
LW-like $A$ & 1 & 3 & $\sigma$ & 
	${m}^2_{\tilde{A}} = \LamEW^2$ & 
	\\
SM-like Z & 1 & 3 & 1 & 
	${m}^2_{Z} = \frac{1}{2} \LamEW^2 \left( 1 - \sqrt{1 - \frac{4 {m}^2_{\hat{Z}}}{\LamEW^2}} \right)$ & 
	\multirow{2}{*}{${m}_{\hat{Z}}^2 = \frac{(g^2+g^{\prime \, 2}) \phi_c^2}{4} $} \\
LW-like Z & 1 & 3 & $\sigma$ & 
	${m}^2_{\tilde{Z}} = \frac{1}{2} \LamEW^2 \left( 1 + \sqrt{1 - \frac{4 {m}^2_{\hat{Z}}}{\LamEW^2}} \right)$ & 
	\\ \hline
SM-like top & $\frac{1}{2}$ & 12 & 1 & 
	${m}_{t}^2 = \frac{2\Lamt^2}{3} \left( 1 - \cos \frac{\theta_t}{3} \right)$ & 
	$\theta_t = \arctan \frac{2 \sqrt{ \alpha ( 1 - \alpha) }}{1 - 2 \alpha}$ \\
LW-like top (1) & $\frac{1}{2}$ &12 & $\sigma$ & 
	${m}_{\tilde{t}_1}^2 = \frac{2\Lamt^2}{3} \left( 1 + \cos \frac{\theta_t + \pi}{3} \right)$ & 
	$\alpha = \frac{27}{4} \frac{{m}_{\hat{t}}^2}{\Lamt^2}$ \\
LW-like top (2) & $\frac{1}{2}$ & 12 & $\sigma$ & 
	${m}_{\tilde{t}_2}^2 = \frac{2\Lamt^2}{3} \left( 1 + \cos \frac{\theta_t - \pi}{3} \right)$ & 
	${m}_{\hat{t}}^2 = h_t^2 \phi_c^2$ \\ \hline
SM-like gluons & 1 & 16 & 1 & 0 & \\ 
LW-like gluons & 1 & 24 & $\sigma$ & $\LamLW^2$ & \\ 
Other SM-like fermions & $\frac{1}{2}$ & 78 & 1 & 0 & \\ 
Other LW-like fermions & $\frac{1}{2}$ & 156 & $\sigma$ & $\LamLW^2$ & \\ 
\hline
\end{tabular}
\caption{\label{tab:LWSM_masses} Tree-level, field-dependent pole
  masses used to construct the LWSM effective potential. $s$, $g$, and
$\sigma$ indicate the spin, effective number of degrees of freedom,
and LW character of the fields; the fifth column gives the mass
eigenvalues in terms of the field-dependent Lagrangian mass parameters
appearing in the last column.}
\end{center}
\end{table}%

Using the results of \sref{sub:Veff_LW_Toy}, the one-loop effective
potential for the LWSM reads\footnote{Here we have neglected the higher-order 
corrections, particularly the so-called ``daisy resummation'' (see, 
{\it e.g.}, \cite{Kapusta:1989, Carrington:1991hz}).  Since, as we discuss below, 
the phase transition is not first order, the daisy resummation does not play a central 
role, in contrast to the case of the Standard Model.  }
\begin{align}\label{eq:Veff_LWSM}
	V_{\rm eff}^{\rm (1L)}(\phi_c, T) & = U(\phi_c) +
\Delta V^{\rm (1L)}_0(\phi_c) + \Delta V^{\rm (1L)}_{T}(\phi_c, T) \ ,
\nn
U(\phi_c) & = \frac{\lambda}{4} \left( \phi_c^2 - v^2 \right)^2 \, ,
\nn
\Delta V_{0}^{\rm (1L)}(\phi_c) & = \delta V_{\rm c.t.} + \sum_i
(-1)^{2s_i} g_i \frac{[m_i^2(\phi_c)]^2}{64 \pi^2} \left[ \ln
m_i^2(\phi_c) - C_{uv} - \frac{3}{2} \right] \, , \nn
\delta V_{\rm c.t.} & = \delta \Omega + \frac{\delta m^2}{2} \phi_c^2
+ \frac{\delta \lambda}{4} \phi_c^4 \, , \nn
	\Delta V_{T}^{\rm (1L)}(\phi_c, T) & = \frac{T^4}{2\pi^2}
        \sum_i \sigma_i g_i \begin{cases} J_B \left(
            \frac{m_i(\phi_c)}{T^2} \right) \, , & s_i = 0 , 1 \, , \\
-J_F
          \left( \frac{m_i^2(\phi_c)}{T^2} \right) \, , & s_i =
          \frac{1}{2} \, ,
\end{cases}
\end{align}
where the sums run over the species listed in \tref{tab:LWSM_masses}.
One calculates $\Delta V_0^{\rm (1L)}$ by evaluating the divergent
momentum integrals in \eref{eq:DeltaV0} using dimensional
regularization ($d = 4-2\epsilon$) and defining the subtraction
constant $C_{UV} \equiv \epsilon^{-1} - \gamma_E + \ln 4 \pi$.  The
bosonic and fermionic thermal functions $J_{B,F}$ are defined in
Eqs.~(\ref{eq:JB_def})~and~(\ref{eq:JF_def}).  We are interested in
the decoupling limit in which the LW mass scale is much greater than
the EW scale.  Since decoupling is not manifest in the $\overline{\rm
MS}$ renormalization scheme, we instead renormalize by requiring that
the tree-level relationships for the Higgs vacuum expectation value and 
mass are maintained
at one loop, and requiring the vanishing of the cosmological constant.
These conditions amount to imposing
\begin{align}
\begin{array}{l}
	0 = \Delta V_0^{\rm (1L)} \Bigr|_{\phi_c = v}
	= \frac{d \Delta V_0^{\rm (1L)}}{d \phi_c} \Bigr|_{\phi_c = v} 
	= \frac{d^2 \Delta V_0^{\rm (1L)}}{d \phi_c^2} \Bigr|_{\phi_c = v} 
\end{array} \com
\end{align}
which may be solved for the counterterms $\delta \Omega$, $\delta
m^2$, and $\delta \lambda$.  After fixing $v = 246 \GeV$, the free
parameters are the four SM couplings $\lambda, g, g^{\prime}, h_t$ and
the five LW mass scales $\LamH, \LamW, \Lamt$, $\LamEW$, and $\LamLW$.
The SM couplings are renormalized to satisfy the tree-level mass
relationships \cite{Beringer:2012zz, Chatrchyan:2012tx, ATLAS:2012ae}
\begin{align}
\begin{array}{lcl}
	\sqrt{ m_{h}^2(v) } = M_H = 125 \GeV \, , & \qquad & \sqrt{
          m_{W^{\pm}}^2(v) } = M_W = 80.4 \GeV \, , \\
	\sqrt{ m_{Z}^2(v) } = M_Z = 91.2 \GeV \, , & \qquad & \sqrt{
          m_{t}^2(v) } = M_t = 172.6 \GeV \, .
\end{array}
\end{align}
For simplicity, we assume a common Lee-Wick mass scale $\LamH = \LamW
= \Lamt = \LamEW = \LamLW$, and take $\LamLW$ as the only free
parameter of the theory.  There is no upper bound on $\LamLW$; the SM
is regained in the limit $\LamLW \gg v$.  Later, we generalize
and discuss phenomenological lower bounds on each of the LW scales.
While most of the LW particles now have masses $<1$ TeV excluded,
 the scalars might be considerably lighter, and therefore for illustration 
we allow $\LamLW$ to be as low as 350 GeV.

\subsection{Finite-Temperature Behavior}\label{sub:FiniteTemp}

\begin{figure}[t]
\hspace{0pt}
\vspace{-0in}
\begin{center}
\includegraphics[width=0.99\textwidth]{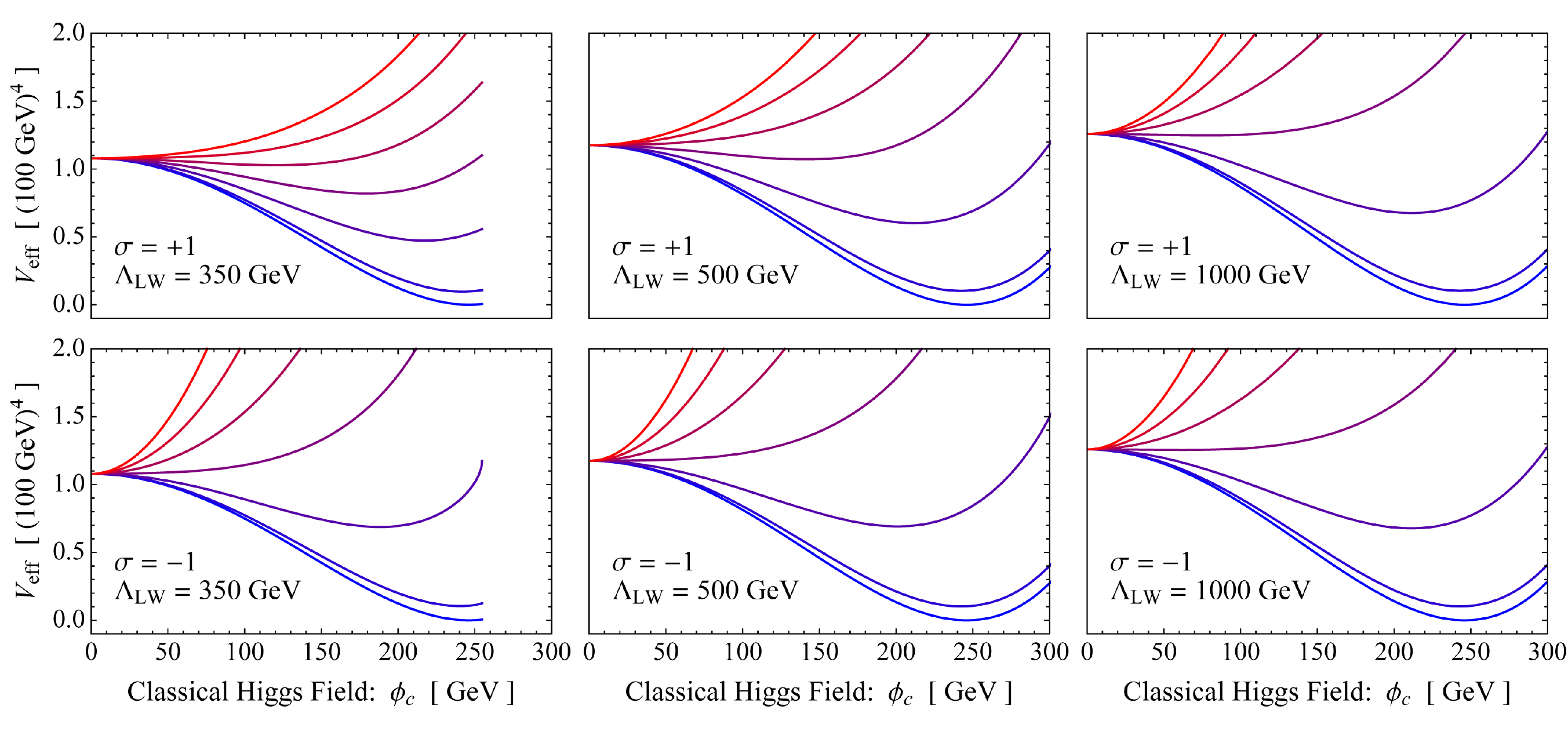} 
\caption{
\label{fig:LWSM_Veff}
Color online.  Variation with respect to temperature $T$ and LW scale $\Lambda_{\rm
  LW}$ of the LWSM thermal effective potential $V_{\rm eff}(\phi_c)$.
$T$ increases from $0 \GeV$ (blue, lowest curves) to $300 \GeV$ (red, highest curves) in increments of
$\Delta T = 50 \GeV$.  }
\end{center}
\end{figure}

\begin{figure}[t]
\hspace{0pt}
\vspace{-0in}
\begin{center}
\includegraphics[width=0.6\textwidth]{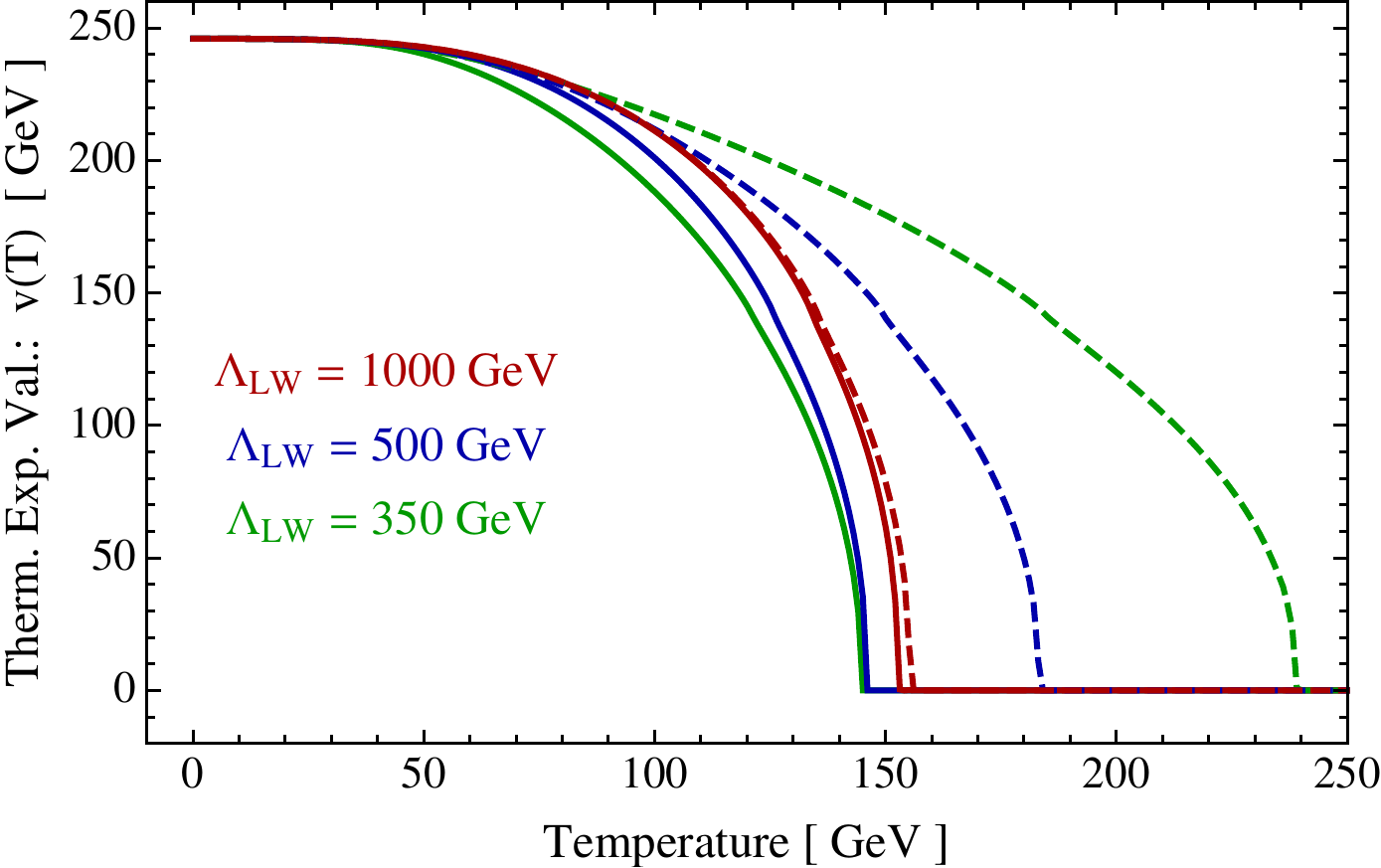} 
\caption{
\label{fig:LWSM_vofT}
Color online.  The electroweak symmetry-breaking order parameter $v(T)$ for the case
of $\sigma = +1$ (dashed) and $\sigma = -1$ (solid).  Note that the pairs of lines for a 
given $\LamLW$ move inward monotonically with increasing $\LamLW$.  The phase
transition temperature is generally higher in the former case due to
the cancellation of the leading $O(T^2)$ temperature dependence
discussed in the text.  }
\end{center}
\end{figure}

The LWSM effective potential $V_{\rm eff}^{\rm (1L)}(\phi_c, T)$ of
\eref{eq:Veff_LWSM} is shown in \fref{fig:LWSM_Veff} over a range of
temperatures and for different values of the LW scale.  In the case
$\LamLW = 350 \GeV$, the curves are not drawn for $\phi_c \ge
\sqrt{4/27} \Lamt / h_t \approx 255 \GeV$, where the LW stability
condition fails in the top sector.  The absence of a barrier in the
effective potential near the critical temperature implies that the
electroweak phase transition is not first order.  This conclusion is
also illustrated in \fref{fig:LWSM_vofT}, where we plot the
electroweak order parameter $v(T)$, {\it i.e.}, the value of $\phi_c$
that minimizes the effective potential, versus $T$.  We define the
phase transition temperature $T_c$ by the condition $v(T
\geq T_c) = 0$.  The absence of a discontinuity in $v(T)$ at $T=T_c$
indicates that the phase transition is not first order.  In this way,
the LWSM electroweak phase transition is similar to the SM phase
transition, and in particular implies that LW electroweak baryogenesis
is not a viable mechanism for the generation of the baryon asymmetry
of the universe \cite{Cohen:1993nk}.

From \fref{fig:LWSM_vofT} one sees that the limit $\LamLW \gg v$
restores the phase transition temperature to $T_c \approx 150 \GeV$,
which corresponds to the SM one-loop result~\cite{Anderson:1991zb}.
In the case $\sigma = +1$ ($-1$), the critical temperature is
generally larger (smaller).  As discussed in \sref{sub:Veff_LW_Toy},
this result can be understood to arise from cancellations between the
positive- and negative-norm fields.  To make this cancellation
explicit in the LWSM, consider the thermal corrections $\Delta
V_T^{\rm (1L)}(\phi_c, T)$ that appear in \eref{eq:Veff_LWSM}.  For
illustrative purposes, one may take the limit $\phi_c \ll \LamLW \ll
T$, in which all species are light compared to the temperature, and
compute
\begin{align}\label{eq:VTexpand}
	\Delta V_T^{\rm (1L)}(\phi_c, T) \approx &
	- \frac{\pi^2}{90} \gast(\sigma) T^4 \nn
	& + T^2 \times \begin{cases}
	+\frac{13}{6} \LamLW^2 & \sigma = +1 \, ,\\
	-\frac{13}{6} \LamLW^2 + \left( \frac{3g^2 + g^{\prime \,
              2}}{16} \! + \! \frac{m_t^2}{2 v^2} \! + \!
          \frac{\lambda}{2} \right) \phi_c^2 + O(\phi_c^4 / \LamLW^2)
        & \sigma = -1 \, , 
	\end{cases} \nn
	& + T \times \begin{cases}
	+\left( \frac{9 g^2 + 3 g^{\prime \, 2} + 3 \lambda}{32 \pi^2}  \right) \phi_c^2 \LamLW + O(\phi_c^3 / \LamLW) & \sigma = +1 \, , \\
	-\left( \frac{9 g^2 + 3 g^{\prime \, 2} + 3 \lambda}{32 \pi^2}  \right) \phi_c^2 \LamLW + O(\phi_c^3 / \LamLW) & \sigma = -1 \, ,
	\end{cases}
\end{align}
where $\gast(\sigma) = 106.75 + 197.5 \sigma$ is a coefficient recognizing that the LW states contribute negatively to the energy density; it should be interpreted as $106.75$ regular and $197.5$ LW degrees of freedom.  
Let us consider each term order-by-order in powers of $T$.

The $O(T^4)$ term has the same form as the free energy density of an
ultra-relativistic gas in which all mass scales are negligible and the
effective number of degrees of freedom is $\gast$.  
Here, the central point of Eqs. (\ref{eq:LWsignflip})-(\ref{eq:LWsamesign}), which propagates through Eqs. (\ref{toybosonenergy})-(\ref{toyfermionenergy}), enters: Each LW degree of freedom contributes a factors $\sigma$ to the potential.  The term
$\gast \ni 106.75$ arises from the SM fields, and the larger term
$\gast \ni 197.5 \sigma$ arises from the LW fields, which outnumber
the SM fields because ($i$) each SM fermion has two LW fermions
[Eq.~(\ref{eq:LWfermiontoy_poles})] and ($ii$) the LW gauge bosons
have explicit masses.  Thus, the sign of $\gast$ follows the sign
of $\sigma$: $\gast(+1) = 304.24$ and $\gast(-1) = -90.75$.  For
the case $\sigma = -1$, we find that the free energy density is
positive, which implies that the pressure, energy density, and entropy
densities are negative.  This result has been obtained previously in
the context of toy LW theories~\cite{Fornal:2009xc,
Bhattacharya:2011bb}, and its interesting implications have been
studied in the context of early universe
cosmology~\cite{Bhattacharya:2012te, Bhattacharya:2013ut}.

In the SM, the term $\Delta V_{T}^{\rm (1L)} \ni T^2 \, {\rm Tr} \,
M^2 \sim T^2 \phi_c^2$ gives rise to symmetry restoration.  In the
$\sigma = +1$ case, this term is absent due to the cancellation
between positive- and negative-norm fields, as already encountered in
\sref{sub:Veff_LW_Toy}.  Then symmetry restoration only comes about
through the subdominant $O(T \LamLW \phi_c^2)$ term, which results in
a higher phase transition temperature.  Note that this term only
arises by virtue of the $O(T)$ nonanalytic term in the expansion of
the {\it bosonic\/} thermal function [see \eref{eq:JB_highT}].
In this case, the fermions are irrelevant for symmetry restoration.

Recently, it has been emphasized \cite{Patel:2011th} that one must take care in extracting gauge-invariant observables from the manifestly gauge-dependent effective potential \cite{Dolan:1973qd}. 
In the conventional phase transition calculation, which we follow here, one obtains $v(T)$ by minimizing $V_{\rm eff}^{(\rm 1L)}$.  
However this definition of the order parameter endows $v(T)$ with the gauge dependence of the effective potential.  
In the case of a first-order phase transition, this gauge dependence can lead to anomalous results for observables such as the baryon number preservation criterion and gravity wave spectrum \cite{Patel:2011th, Wainwright:2011qy, Wainwright:2012zn}.  
However, \cite{Garny:2012cg} have pointed out that the gauge dependence is small as long as the perturbative expansion is valid.  
Since the LWSM phase transition is not first order, the only potentially gauge-dependent parameter is $T_c$.  
To check our results, we have also calculated the critical temperature using the technique of \cite{Patel:2011th}.  
We find qualitative agreement with the critical temperatures presented in \fref{fig:LWSM_vofT}, namely, that $T_c$ is generally increased (decreased) for the case $\sigma = +1$ ($-1$) with respect to the SM value, but the gauge-invariant $T_c$ is typically O(20-35\%) smaller, with the discrepancy at the larger end of this range at lower $\LamLW$.  
The preceding discussion of the unusual high-temperature behavior of the LWSM is independent of this gauge-fixing ambiguity because the symmetric phase, $\phi_c = 0$, is a critical point of the effective potential.  

For simplicity we have assumed a common LW mass scale $\LamLW \geq 350
\GeV$ up to this point, but let us now discuss a more
phenomenologically motivated parameter set.  The strongest constraints
on the mass of the LW partners come from electroweak precision tests.
The oblique parameter $T$ is sensitive to the LW top, and its
constraints impose\footnote{These bounds are derived assuming a Higgs
mass $M_H = 115 \GeV$, which deviates by $O(10\%)$ from the value $M_H
\approx 125 \GeV$ subsequently measured by the LHC\@.  This shift
translates into a comparable shift in the bounds, which is
insignificant to the level of precision with which we have been
working.  } $\Lamt > 1.5 \TeV$ at $95\%$ CL \cite{Chivukula:2010nw}.
The oblique parameters $W$ and $Y$ are sensitive to the LW gauge
bosons and impose $\LamW = \LamB > 2.3 \TeV$ at $95\%$ CL
\cite{Chivukula:2010nw}.  Finally, the $Z \to b \bar{b}$ branching
fraction and forward-backward asymmetry are sensitive to the new
charged scalars and impose $\LamH > 640 \GeV$ at $95 \%$ CL
\cite{Lebed:2012ab} (see also \cite{Carone:2009nu}).  Setting each of
these parameters to its lower bound, we calculate the electroweak
order parameter $v(T)$ and find it to be indistinguishable from the
solid red ({\it i.e.}, innermost) curve of \fref{fig:LWSM_vofT}.  The phase transition
temperature is $T_c \approx 150 \GeV$, which is very close to the SM
one-loop result ({\it i.e.}, the $\LamLW \to \infty$ limit).  
Note that this result is obtained despite the weaker
bound on the LW Higgs mass, which presumably could have played a
significant role; it suggests that the departures from the SM phase
transition seen in \fref{fig:LWSM_vofT} are driven primarily by the LW
tops, which at $350 \GeV$ are near the threshold of their LW stability
condition.

\section{Conclusions}\label{sec:Conclusions}

This paper has two goals: to explore the thermodynamics of LW theories
and to study the LWSM at temperatures $T \gtrsim O(100-1000 \GeV)$.
In a LW theory, the negative-norm degrees of freedom are forbidden
from appearing as external states in S-matrix elements by the LW/CLOP
prescription, but it is unclear how to implement these constraints
when the system is brought to finite temperature.  If no special
consideration is paid to the negative-norm degrees of freedom, then at
leading order in the interactions these LW particles can be treated as
a free (ideal) gas, and the partition function is calculated in the
standard way.  However, it seems that this picture is incompatible
with the LW/CLOP prescription.  Alternatively, the negative-norm
particles can be restricted to internal lines, where they simply
modify the energy of the states containing positive-norm SM particles.
In the limit of small couplings the LW particles become narrow
resonances, but since their decay width is negative (a consequence of
the negative-residue propagator), their contribution to the
thermodynamic variables is just the opposite of what one would expect
for an ideal gas of SM particles.  Since some uncertainty remains as
to the correct sign of the LW particle contribution to the free energy
compared to that of its SM partner, we introduce the index $\sigma =
\pm 1$ to consider both cases.

We find that the LWSM electroweak phase transition is qualitatively
very similar to the Standard Model crossover.  For phenomenologically
viable values of the LW mass scale $\LamLW = O({\rm TeV})$, the LW
degrees of freedom are heavy and decouple from the physics of the
electroweak phase transition that occurs at $T = O(100 \GeV)$.
However, at temperatures comparable to $\LamLW$, the LW fields yield
significant modifications to the thermodynamics.  One finds in case
$\sigma = +1$ a cancellation of the leading $O(T^2)$ correction to the
Higgs mass.  Since this effective mass is responsible for symmetry
restoration, the cancellation tends to retard symmetry restoration and
increases the phase transition temperature.  In the ultrarelativistic
limit, the number of effective species is found to be $\gast = (106.75
+ 197.5 \sigma)$, where the first term is the standard SM contribution
and the second term arises from their LW partners, which are greater
in number because of the doubling in the fermion sector and explicit
gauge boson partner masses.  In the case $\sigma = -1$, the LW
partners overwhelm the SM degrees of freedom to give $\gast < 0$,
implying a negative pressure and energy density.

Our results have immediate implications for early-universe cosmology.
Since the electroweak phase transition is not first order, LW
electroweak baryogenesis is not a viable explanation for the baryon
asymmetry of the universe, nor do we expect other cosmological relics,
such as gravitational waves, to be produced.
References~\cite{Bhattacharya:2012te, Bhattacharya:2013ut} studied the
effect of LW theories on very early universe cosmology using the case
that we call $\sigma = -1$.  They find that the unusual thermodynamic
properties of LW theories can lead to novel features, such as bouncing
cosmologies and mini-reheating events when LW particles decouple.
Alternately, if LW thermodynamics is correctly described by the case
$\sigma = +1$, then the early-universe cosmology very much resembles
the SM concordance model, but with the addition of $197.5$
relativistic degrees of freedom.

An interesting generalization of our work would be to consider the
$N=3$ LWSM~\cite{Carone:2008iw}.  In that extension, the
phenomenological bounds on the LW scale are weaker, and the LW
partners may have a more significant impact on the nature of the phase
transition.  Moreover, many additional degrees of freedom contribute
to $\gast$ and thereby manifest themselves in the high-temperature
thermodynamics.  Finally, if the narrow-resonance approximation is not
valid for some of the LW particles in the LWSM (for any $N$), then a
more careful analysis than presented here is required.

\begin{acknowledgments}
  This work was supported by the National Science Foundation under
  Grant No.\ PHY-1068286 (RFL and RHT) and by the Department of Energy
  under Grant No.\ DE-SC0008016 (AJL).
\end{acknowledgments}

\begin{appendix}

\section{Quantization Conventions} \label{sec:quant_conv}

\subsection{Classical to Quantum Theory}
A first, essential step to performing calculations with negative-norm
states is the establishment of consistent conventions for
quantization, time ordering, and the Feynman rules.  Since so much
potential confusion can arise from improperly handled sign
conventions, we begin with the pedagogical exercise of presenting
textbook expressions, augmented with the relevant signs.  Suppose
first that one is given the classical Lagrangian density
\begin{equation}
{\cal L} = \eta_H \left( \frac{1}{2}\dot{\phi}^2 -
  \frac{1}{2}m^2\phi^2 \right) \, , \label{classicallagrangian}
\end{equation}
from which one sees that $\pi_{\phi} \equiv \partial{\cal
  L}/\partial{\dot{\phi}}= \eta_H \dot{\phi}$, and therefore
\begin{equation}
{\cal H} = \eta_H \left( \frac{1}{2}\pi_{\phi}^2 +
  \frac{1}{2}m^2\phi^2 \right) \, . \label{classicalhamiltonian}
\end{equation}
The sign $\eta_H$ is therefore defined so that $\eta_H = \pm 1$ gives
a semipositive(negative)-definite Hamiltonian density.  In order to
quantize this theory, one must impose quantization conditions on the
fields and their conjugate momenta; however, the sign $\eta_C$ of the
fundamental commutation relation may be allowed to vary while still
allowing unitary time evolution:
\begin{equation}
[\phi (\mathbf{x}) ,\,\pi_{\phi} (\mathbf{y})]
=i\eta_C \delta^{(3)} (\mathbf{x} - \mathbf{y})
\label{canonicalcom} \, .
\end{equation}
How does this choice affect the time evolution of states defined on a
Hilbert space?  For the fundamental field operators
$\phi,\,\pi_{\phi}$,
\begin{eqnarray}
\left[ \phi, \, H \right] & = & \eta_H \left[ \phi(\mathbf{x}), \,
\frac{1}{2} \int \! d^3 \! y \, \pi_{\phi}^2 (\mathbf{y}) \right]
= i \eta_H \eta_C \pi_{\phi} 
= i \eta_C \dot{\phi} \, , \nonumber \\
\left[\pi_{\phi} \, , H \right] & = & \eta_H
\left[ \pi_{\phi} (\mathbf{x}), \, \frac{1}{2} \, m^2 \! \! \int d^3
\! y \, \phi^2 (\mathbf{y}) \right]
= -i \eta_H \eta_C m^2 \phi
= i\eta_C \dot{\pi}_{\phi} \, ,
\label{phipiH}
\end{eqnarray}
which uses the definition of $\pi_\phi$, the commutation relation
Eq.~(\ref{canonicalcom}), and the commutator identity $[A,BC] = [A,B]C
+ B[A,C]$, while the final equality also uses the Hamilton equation of
motion, $\dot{\pi}_{\phi}=-\partial{\cal H}/\partial{\phi} = -\eta_H
m^2 \phi$.  From the above relations, one may prove the more general
Heisenberg equation:
\begin{equation}
[\mathcal{O},\,H]=i\eta_C\dot{\mathcal{O}}\label{hamcom} \, ,
\end{equation}
for any function $\mathcal{O} (\phi, \pi_{\phi})$.  The proof is
straightforward: Both sides of Eq.~(\ref{hamcom}) are linear in
$\mathcal{O}$, so without loss of generality $\mathcal{O}$ may be
taken as a monomial in $\phi$ and $\pi_{\phi}$.  The commutator
identity $[AB,H]=A[B,H]+[A,H]B$ shows that, if $A$ and $B$ separately
satisfy Eq.~(\ref{hamcom}), then the right-hand side is $A \dot{B} + B
\dot{A} = \dot{(AB)}$, which means that $AB$ also satisfies
Eq.~(\ref{hamcom}).  Since Eqs.~(\ref{phipiH}) show that $\phi$ and
$\pi_{\phi}$ themselves satisfy Eq.~(\ref{hamcom}), then by induction
so does any arbitrarily complicated function $\mathcal{O}$ of them.
These results indicate that, once the phase space of the system is
partitioned into one set for which $\eta_C=1$ in
Eq.~(\ref{canonicalcom}) and another set for which $\eta_C=-1$, the
operators defined in those partitions obey separate Heisenberg
equations of motion parameterized by $\eta_C$ (and independent of
$\eta_H$).  We do not consider operators that are functions of fields
or their conjugate momenta drawn from both partitions.

Now, how does one interpret the potentially ``wrong-sign'' Heisenberg
equations of motion in Eq.~(\ref{hamcom})?  They may be exponentiated
to obtain
\begin{equation}
  e^{i\eta_C H(t-t_0)}\mathcal{O}(t_0)e^{-i\eta_C H(t-t_0)}
  =\mathcal{O}(t) \, . \label{timedep}
\end{equation}
The fields are therefore evolved forward in time from $t_0$ to $t$ by
means of the unitary operator $U(t,t_0) \equiv \mathrm{exp}[-i\eta_C
H(t-t_0)]$.  For the $\eta_C=-1$ partition, this operator has the
opposite phase compared to the conventional time evolution operator
encountered in quantum field theory.  We show next that the choice of
$\eta_C$ has crucial implications for quantities such as the Feynman
propagator of the Lee-Wick field theories, as well as the allowed
direction of Wick rotation in the complex $p^0$ plane.

\subsection{Mode Expansions and the Hamiltonian}

We begin with the mode expansions for a Lee-Wick type field
$\phi(\mathbf{x})$ and its conjugate momentum $\pi_\phi(\mathbf{x})$:
\begin{eqnarray}
  \phi(\mathbf{x}) & = &
  \int\frac{d^3p}{(2\pi)^3}\frac{1}{\sqrt{2\omega_p}}
  \left(a_{\mathbf{p}}^{\phantom\dagger}e^{i\mathbf{p\cdot x}}
    +a_{\mathbf{p}}^{\dagger}e^{-i\mathbf{p\cdot x}}\right)=
  \int\frac{d^3p}{(2\pi)^3}\frac{1}{\sqrt{2\omega_p}}
  \left(a_{\mathbf{p}}^{\phantom\dagger}+a_{\mathbf{-p}}^{\dagger}
  \right)e^{i\mathbf{p\cdot x}} \, , \label{phimodes} \\
  \pi_\phi(\mathbf{x}) & = & \eta_H \int\frac{d^3p}{(2\pi)^3}(-i)
  \sqrt{\frac{\omega_p}{2}}\left(a_{\mathbf{p}}^{\phantom\dagger}
    e^{i\mathbf{p\cdot
        x}}-a_{\mathbf{p}}^{\dagger}e^{-i\mathbf{p\cdot
        x}}\right) \nonumber \\ & = &
\eta_H \int\frac{d^3p}{(2\pi)^3}(-i)
  \sqrt{\frac{\omega_p}{2}}\left(a_{\mathbf{p}}^{\phantom\dagger}
    -a_{\mathbf{-p}}^{\dagger}\right)e^{i\mathbf{p\cdot
      x}} \, ,
\end{eqnarray}
where $\omega_p \equiv +\sqrt{\mathbf{p}^2+m^2}$ is strictly positive,
and the factor $\eta_H$ reflects the result $\pi_{\phi} = \eta_H
\dot{\phi}$.  The canonical commutation relation
$[\phi(\mathbf{x}),\pi(\mathbf{y})]=i\eta_C\delta^{(3)}(\mathbf{x-y})$
constrains the commutator
$[a_{\mathbf{p}}^{\phantom\dagger},a_{\mathbf{q}}^{\dagger}]\equiv
  \eta_N (2\pi)^3\delta^{(3)} (\mathbf{p-q})$:
\begin{eqnarray}
  [\phi(\mathbf{x}),\pi(\mathbf{y})] & = & -\eta_H \int\frac{d^3p\,d^3q}
  {(2\pi)^6}\frac{i}{2}\sqrt{\frac{\omega_q}{\omega_p}}
  \left([a_\mathbf{-p}^{\dagger},a_{\mathbf{q}}^{\phantom\dagger}]
    -[a_{\mathbf{p}}^{\phantom\dagger},a_{\mathbf{-q}}^{\dagger}]\right)
  e^{i\mathbf{(p\cdot x+q\cdot y)}} \nonumber \\
  & = & \eta_H \int\frac{d^3p\,d^3q}{(2\pi)^6}\frac{i}{2}
  \sqrt{\frac{\omega_q}{\omega_p}} \, 2(2\pi)^3\eta_N\delta^{(3)}
  (\mathbf{p+q})e^{i(\mathbf{p\cdot x+q\cdot y)}}
\nonumber \\ & = & i\eta_H \eta_N
\delta^{(3)}(\mathbf{x-y}) \,
  , \label{canonicalcom2}
\end{eqnarray}
from which one identifies $\eta_C= \eta_H \eta_N$.  The interpretation
of $\eta_N$ becomes apparent once one calculates the spectrum of the
theory.  The mode expansion of the Hamiltonian reads
\begin{eqnarray}
  H & = & \eta_H \int d^3x\left[\frac{1}{2}\pi^2+\frac{1}{2}(\nabla
    \phi)^2+\frac{1}{2}m^2\phi^2\right] \nonumber \\
  & = & \eta_H \frac{1}{2}\int d^3x\int\frac{d^3 \! p \, \, d^3 \! q}
  {(2\pi)^6} e^{i(\mathbf{p+q)\cdot x}}
\nonumber \\ & & 
  \cdot \left[-\frac{\sqrt{\omega_p\omega_q}}{2}
    (a_{\mathbf{p}}^{\phantom\dagger}-a_{-\mathbf{p}}^{\dagger})
    (a_{\mathbf{q}}^{\phantom\dagger}-a_{-\mathbf{q}}^{\dagger})
    +\frac{-\mathbf{p\cdot q}+m^2}{2\sqrt{\omega_p\omega_q}}
    (a_{\mathbf{p}}^{\phantom\dagger}+a_{-\mathbf{p}}^{\dagger})
    (a_{\mathbf{q}}^{\phantom\dagger}+a_{-\mathbf{q}}^{\dagger})\right]
  \nonumber \\
  & = & \eta_H \displaystyle\int\frac{d^3p}{(2\pi)^3} \, \omega_p
  \left( a_{\mathbf{p}}^{\dagger}a_{\mathbf{p}}^{\phantom\dagger} +
    \frac 1 2 \left[ a_{\mathbf{p}},  a_{\mathbf{p}}^{\dagger} \right]
  \right) \, \, .
\end{eqnarray}
From this expansion follows the commutators:
\begin{eqnarray}
\left[ H,a_{\mathbf{p}}^{\phantom\dagger} \right] & = & \eta_H
\int\frac{d^3q} {(2\pi)^3} \,
\omega_{q}[a_{\mathbf{q}}^{\dagger},a_{\mathbf{p}}^{\phantom\dagger}]
\, a_{\mathbf{q}}^{\phantom\dagger}= -\eta_H \eta_N \, \omega_p \,
a_{\mathbf{p}}^{\phantom\dagger} = -\eta_C \, \omega_p \,
a_{\mathbf{p}}^{\phantom\dagger} \, , \label{lowercom} \\
\left[ H,a_{\mathbf{p}}^{\dagger} \right] & = & \eta_H
\int\frac{d^3q}{(2\pi)^3} \, \omega_{q} \, a_{\mathbf{q}}^{\dagger}
[a_{\mathbf{q}}^{\phantom\dagger}, a_{\mathbf{p}}^{\dagger}]
= +\eta_H \eta_N \, \omega_p \, a_{\mathbf{p}}^{\dagger}
= +\eta_C \, \omega_p \, a_{\mathbf{p}}^{\dagger} \, .
\label{raisingcom}
\end{eqnarray}
These commutation relations immediately provide the time dependence of
the ladder operators.  Rearranging Eq.~(\ref{lowercom}) into
$Ha_{\mathbf{p}}^{\phantom\dagger}=a_{\mathbf{p}}^{\phantom\dagger}
(H-\eta_C\omega_p)$, one acts repeatedly with $H$ on the left to
obtain $H^na_{\mathbf{p}}^{\phantom\dagger}=
a_{\mathbf{p}}^{\phantom\dagger}(H-\eta_C\omega_p)^n$, which may be
exponentiated [consistent with Eq.~(\ref{timedep})] to
\begin{eqnarray}
a_{\mathbf{p}}^{\phantom\dagger}(t) & = &
U^{\dagger}(t,0)a_{\mathbf{p}}^{\phantom\dagger}(t=0)U(t,0)
=e^{i\eta_C Ht}a_{\mathbf{p}}^{\phantom\dagger}e^{-i\eta_C Ht}
=a_{\mathbf{p}}^{\phantom\dagger}e^{-i\omega_p t} \, .
\end{eqnarray}
Hermitian conjugation of this result immediately gives
$a_{\mathbf{p}}^{\dagger}(t)=a_{\mathbf{p}}^{\dagger} \, \mathrm{exp}
(i\omega_p t)$.  In the case $\eta_H = -1$,
Eqs.~(\ref{lowercom})--(\ref{raisingcom}) show that the choice $\eta_N
= -1$ still leads to $a_{\mathbf{p}}^{\dagger}$ and
$a_{\mathbf{p}}^{\phantom\dagger}$ acting as raising and lowering
operators, respectively.  One may then define a lowest-energy state
$|0\rangle$ such that $a_{\mathbf{p}}^{\phantom\dagger}|0\rangle=0$,
with single-particle momentum eigenstates
$|\mathbf{p}\rangle=\sqrt{2\omega_p} \,
a_{\mathbf{p}}^{\dagger}|0\rangle$ whose norms are given by
\begin{equation}
\langle\mathbf{p}|\mathbf{q}\rangle=2\sqrt{\omega_p\omega_q}
\langle 0|a_{\mathbf{p}}^{\phantom\dagger}
a_{\mathbf{q}}^{\dagger}|0\rangle=2\sqrt{\omega_p\omega_q}
\langle 0|[a_{\mathbf{p}}^{\phantom\dagger},a_{\mathbf{q}}^{\dagger}]
|0\rangle=2\eta_N\omega_p \, \delta^{(3)}(\mathbf{p-q}) \, ,
\label{normdef}
\end{equation}
from which one concludes that the $\eta_N=-1$ convention corresponds
to states of negative norm.  In short,
\begin{itemize}
\item $\eta_H$ is defined by the Hamiltonian (or Lagrangian), \eqref{classicallagrangian}-\eqref{classicalhamiltonian};
\item $\eta_C$ is defined by the $[\phi,\,\pi]$ commutation relation \eqref{canonicalcom};
\item $\eta_N$ is defined by the $[a^{\phantom\dagger},\,a^{\dagger}]$ commutation relation (above \eqref{canonicalcom2}) and fixes the norm.
\end{itemize}

\noindent If instead, $\eta_N=1$, the spectrum is
defined by raising and lowering operators
$a_{\mathbf{p}}^{\phantom\dagger}$ and $a_{\mathbf{p}}^{\dagger}$,
respectively, and one must then define the vacuum in a sensible way.
It is possible to choose $a_{\mathbf{p}}^{\dagger}|0\rangle=0$ and
build successive $n$-particle states with repeated action of
$a_{\mathbf{p}}^{\phantom\dagger}$, but this choice effectively
amounts just to exchanging the roles of 
$a_{\mathbf{p}}^{\dagger}$ and
$a_{\mathbf{p}}^{\phantom\dagger}$.  We instead choose the
prescription of defining a {\it highest}-energy state such
that $a_{\mathbf{p}}^{\phantom\dagger}|0\rangle=0$, and create
negative-energy modes using
$|\mathbf{p}\rangle=\sqrt{2\omega_p}a_{\mathbf{p}}^{\dagger}|0\rangle$.
The Hamiltonian spectrum (ignoring the zero-point energy) for either
sign of $\eta_H$ then becomes
\begin{eqnarray}
  H|\mathbf{p}\rangle & = & \left(\eta_H \int \! \frac{d^3q}{(2\pi)^3}
\, \omega_q \, a_{\mathbf{q}}^{\dagger}
a_{\mathbf{q}}^{\phantom\dagger}\right)a_{\mathbf{p}}^{\dagger}
\sqrt{2\omega_p}|0\rangle= \eta_H \int\frac{d^3q}{(2\pi)^3} \, \omega_q
a_{\mathbf{q}}^{\dagger}[a_{\mathbf{q}}^{\phantom\dagger},
a_{\mathbf{p}}^{\dagger}]\sqrt{2\omega_p}|0\rangle\nonumber \\
  & = & \eta_H \eta_N\omega_pa_{\mathbf{p}}^{\dagger}\sqrt{2\omega_p}
|0\rangle= \eta_H \eta_N\omega_p| \, \mathbf{p}\rangle \nonumber \, , \\
  \therefore H|\mathbf{p}\rangle & \equiv & 
  E_p|\mathbf{p}\rangle = \eta_H \eta_N\omega_p| \, \mathbf{p}\rangle
= \eta_C \omega_p| \, \mathbf{p} \rangle \, . \label{hamspec}
\end{eqnarray}
We see that, working in a convention in which there exists a state
$\left| 0 \right>$ annihilated by the operators
$a_{\mathbf{p}}^{\phantom\dagger}$, the sign $\eta_C$ defined in
Eq.~(\ref{canonicalcom}) uniquely determines the sign of the energy
eigenvalues.  The choice $\eta_H = \eta_C = \eta_N = +1$ is of course
the conventional Klein-Gordon theory with positive energies and
positive norms, whereas one-particle states in Lee-Wick theories
($\eta_H = -1$) can have either negative norms ($\eta_N = -1$) and
positive energies ($\eta_C = +1$) [the conventional formulation] or
positive norms ($\eta_N = +1$) and negative energies ($\eta_C = -1$).

\subsection{Calculating the Propagator}

For the zero-temperature quantum theory, it is important to define the
Feynman propagator so that calculations may be performed in a
straightforward manner; to do so relies on how one performs contour
integrals in the complex $p^0$ plane.  This discussion ultimately
leads to a proper choice of the $\pm i\epsilon$ terms in the
propagator, as well as singling out the unique Wick rotation allowed
in defining loop integrals.  To begin with, one writes down a
mode expansion for the fields generalizing Eq.~(\ref{phimodes}):
\begin{equation}
\phi(x)=\int\frac{d^3p}{(2\pi)^3}\frac{1}{\sqrt{2\omega_p}}
\left(a_{\eta_C \mathbf{p}}^{\phantom\dagger}e^{-ip\cdot x}+
a_{\eta_C \mathbf{p}}^{\dagger}e^{ip\cdot x}\right)\bigg|_{\eta_C}
\label{scalarfield} \, ,
\end{equation}
where the $\eta_C$ evaluation is shorthand for $p^0$ evaluation at
$p^0=\eta_C\omega_p = \eta_H \eta_N \omega_p$.  This bifurcation
admits the possibility of defining the field on either the positive-
or negative-mass shell ({\it i.e.},
$p^0=\pm\sqrt{\mathbf{p}^2+m^2}$).  The Lorentz invariance of $p \cdot
x$ in Eq.~(\ref{scalarfield}) must be maintained in either case, and
so we generalize the ladder operators to create and destroy states of
momentum $\eta_C\mathbf{p}$.

The first step in calculating the Feynman propagator is to obtain the
two-point function, $\langle 0|\phi(x)\phi(y)|0\rangle$.  For
arbitrary $\eta_N$, the single non-vanishing term of the transition
amplitude is
\begin{eqnarray}
\label{prop}
\langle 0|\phi(x)\phi(y)|0\rangle & = &
\int\frac{d^3 \! p \, d^3 \! q}{(2\pi)^6}
\frac{1}{2\sqrt{\omega_p\omega_q}}
\langle0|a_{\eta_C\mathbf{p}}^{\phantom\dagger}
a_{\eta_C\mathbf{q}}^{\dagger}|0\rangle
e^{-i(p\cdot x-q\cdot y)}\big|_{\eta_C}\nonumber \\
& = &
\int\frac{d^3 \! p \, d^3 \! q}{(2\pi)^3}\frac{\eta_N}
{2\sqrt{\omega_p\omega_q}}
\delta^{(3)}[\eta_C(\mathbf{p-q})]e^{-i(p\cdot
  x-q\cdot
  y)}\big|_{\eta_C} \nonumber \\ & = & \int\frac{d^3p}{(2\pi)^3}
\frac{\eta_N}{2\omega_p}e^{-ip\cdot(x-y)}\big|_{\eta_C} \nonumber \\
& \equiv & D^{\eta}(x-y) \, .
\end{eqnarray}
The superscript $\eta$ serves as a bookkeeping tool to remember which
quantization scheme one is using.  Now that the form of the two-point
function is determined, one constructs the time-ordered Feynman
propagator:
\begin{eqnarray}
D_F^{\eta} (x-y) & \equiv & \theta(x^0-y^0)D^{\eta}(x-y)
+\theta(y^0-x^0) D^{\eta}(y-x) \nonumber \\
& = & \int\frac{d^3p}{(2\pi)^3}\frac{\eta_N}{2\omega_p}
\left[\theta(x^0-y^0)e^{-ip\cdot(x-y)}\big|_{p^0=\eta_C\omega_p}
+\theta(y^0-x^0)e^{ip\cdot(x-y)}\big|_{p^0=\eta_C\omega_p}\right]
\nonumber \\
& = & \int\frac{d^3p}{(2\pi)^3}\frac{\eta_N}{2\omega_p}
\left[\theta(x^0-y^0)e^{-ip\cdot(x-y)}\big|_{p^0=\eta_C\omega_p}
+\theta(y^0-x^0)e^{-ip\cdot(x-y)}\big|_{p^0=-\eta_C\omega_p}\right]
\nonumber \\
& = & \int\frac{d^4p}{(2\pi)^3}\frac{\eta_N}{2\omega_p}
\left[\theta(x^0-y^0)\delta(p^0 \! -\eta_C\omega_p)+\theta(y^0-x^0)
\delta(p^0 \! +\eta_C\omega_p)\right]e^{-ip\cdot(x-y)} .
\label{feynprop1}
\end{eqnarray}
To continue, we invoke the Lee-Wick prescription: The theory must be
free of exponentially growing outgoing modes.  This condition
determines how the poles are to be pushed above and below the real
$p^0$ axis as a function of the parameter $\eta_C$.
Equation~(\ref{feynprop1}) may now be rewritten as
\begin{eqnarray}
& & \int\frac{d^4p}{(2\pi)}\frac{\eta_N}{2\omega_p}
[\delta(p^0-\eta_C\omega_p+i\epsilon)+
\delta(p^0+\eta_C\omega_p-i\epsilon)]e^{-ip\cdot(x-y)} \nonumber \\
& = & \int\frac{d^4p}{(2\pi)^3}\frac{\eta_N}{2\omega_p}
\left(\frac{1}{-2\pi i} \cdot \frac{1}{p^0-(\eta_C-i\epsilon)}+
\frac{1}{2\pi i} \cdot \frac{1}{p^0+(\eta_C\omega_p-i\epsilon)}\right)
e^{-ip\cdot(x-y)} \nonumber \\
& = & \int\frac{d^4p}{(2\pi)^4}\frac{i\eta_N}{2\omega_p}
\left(\frac{p^0+\eta_C\omega_p-i\epsilon-p^0+\eta_C
\omega_p-i\epsilon}{(p^0)^2-(\eta_C\omega_p-i\epsilon)^2}\right)
e^{-ip\cdot(x-y)}\notag\\
& = & \int\frac{d^4p}{(2\pi)^4}\frac{i\eta_N\eta_C}
{p^2-m^2+i\eta_C\epsilon}e^{-ip\cdot(x-y)},
\end{eqnarray}
from which one obtains the momentum-space Feynman propagator
\begin{equation}
\tilde{D}^{\eta}_F(p)=\frac{i\eta_H}{p^2-m^2+i\eta_C\epsilon} \, .
\label{feynprop2}
\end{equation}
Examining the structure of Eq.~(\ref{feynprop2}), one sees that
Lee-Wick theories of either quantization possess the hallmark
``wrong-sign'' propagator, since $\eta_H = \eta_C\eta_N=-1$ for them.
The conventional Klein-Gordon propagator may also be recovered upon
setting $\eta_H=\eta_C=\eta_N=+1$.  However, one subtlety does exist
for the case $\eta_C=-\eta_N=1$: The Feynman prescription for
integrating around the poles has the opposite sign with respect to the
usual case.  This means that the shifted poles lie in the first and
third quadrants, rather than the fourth and second; therefore, when
one attempts a Wick rotation upon evaluating a loop integral, the
proper substitution is $p_0=-ip^0_E$, corresponding to
{\it counterclockwise} rotation in the complex $p^0$ plane.

\section{The LWSM Spectrum}\label{sec:Derive_Masses}

For completeness, we present here the calculation of the
field-dependent masses that appear in \tref{tab:LWSM_masses}.  We use
the metric convention $g_{\mu \nu} = {\rm diag}(1,-1,-1,-1)$.

\subsection{Higgs \& Electroweak Gauge Sector}\label{sub:LWSM_Higgs}

In the higher-derivative formalism, we denote the Higgs doublet as
$\hat{H}$, the $\SU{2}_L$ gauge field as $\hat{W}^{a}_{\mu}$, and the
$\U{1}_Y$ gauge field as $\hat{B}_{\mu}$.  We suppose that there is a
nonzero homogenous Higgs condensate $\langle \hat{H} \rangle = ( 0 \,
, \, \phi_c / \sqrt{2} )^T$ that breaks the electroweak symmetry down
to $\U{1}_{\rm EM}$.  The Higgs field may be expanded about the
background as
\begin{align}\label{eq:H_expansion}
	\hat{H} = \begin{pmatrix} \hat{h}^+ \\ \frac{ \phi_c + \hat{h}
+ i \hat{P} }{\sqrt{2}} \end{pmatrix} \, ,
\end{align}
where $\hat{h}$ and $\hat{P}$ are real scalar fields and $\hat{h}^+$
is complex.  After electroweak symmetry breaking, we denote the
photon, neutral weak boson, and charged weak boson fields as
$\hat{A}_{\mu}$, $\hat{Z}_{\mu}$, and $\hat{W}^{\pm}_{\mu}$
respectively.  These are related to the original electroweak gauge
fields by the standard transformations
\begin{align}\label{eq:ZAW_rotation}
\begin{array}{lll}
	\hat{Z}_{\mu} &=& \cos \theta_W \, \hat{W}^3_{\mu} - \sin
\theta_W \, \hat{B}_{\mu} \, , \\
	\hat{A}_{\mu} &=& \sin \theta_W \, \hat{W}^3_{\mu} + \cos
\theta_W \, \hat{B}_{\mu} \, , \\
	\hat{W}^{\pm}_{\mu} &=& \frac{1}{\sqrt{2}} \left(
\hat{W}^1_{\mu} \mp i \hat{W}^2_{\mu} \right) \, ,
\end{array}
\end{align}
where $\cos \theta_W = g / \sqrt{g^2 + g^{\prime \, 2}}$ and $\sin
\theta_W = g^{\prime} / \sqrt{g^2 + g^{\prime \, 2}}$.  We work in the
$R_{\xi}$ gauge formalism for generality and restrict to the Landau
gauge ($\xi=0$) at the end.  We introduce eight anti-commuting, scalar
ghost fields $c_A$, $c_Z$, $c_{W^+}$, $c_{W^-}$, $\bar{c}_A$,
$\bar{c}_Z$, $\bar{c}_{W^+}$, and $\bar{c}_{W^-}$.

The gauge-fixed LWSM electroweak sector is specified by the Lagrangian
\begin{align}\label{eq:LWSM_Higgs}
	&\mathcal{L}_{\rm hd}^{\rm (EW)} = \mathcal{L}_{\rm hd}^{\rm
(H)} + \mathcal{L}_{\rm hd}^{\rm (B)} + \mathcal{L}_{\rm hd}^{\rm (W)}
+ \mathcal{L}_{\rm hd}^{\rm (g.f.)} + \mathcal{L}_{\rm hd}^{\rm (gh.)}
\, , \\
	&\qquad \begin{array}{l}
	\mathcal{L}_{\rm hd}^{\rm (H)} = \abs{ \hat{D}_{\mu} \hat{H}
}^2 - \frac{1}{\LamH^2} \abs{ \hat{D}_{\mu} \hat{D}^{\mu} \hat{H} }^2
- U_{\rm hd}(\hat{H}) \, , \\
	\mathcal{L}_{\rm hd}^{\rm (B)} = - \frac{1}{4} \hat{B}_{\mu
\nu} \hat{B}^{\mu \nu} + \frac{1}{2 \LamB^2} \left( \partial^{\mu}
\hat{B}_{\mu \nu} \right)^2 \, , \\
	\mathcal{L}_{\rm hd}^{\rm (W)} = - \frac{1}{4}
\hat{W}^{a}_{\mu \nu} \hat{W}^{a \, \mu \nu} + \frac{1}{2 \LamW^2}
\left( D^{\mu} \hat{W}^{a}_{\mu \nu} \right)^2 \, , \\
	\mathcal{L}_{\rm hd}^{\rm (g.f.)} = - \frac{1}{2 \xi_A} \left(
\partial^{\mu} \hat{A}_{\mu} \right)^2 - \frac{1}{2 \xi_Z} \left(
\partial^{\mu} \hat{Z}_{\mu} - \xi_Z \, \frac{\sqrt{g^2 + g^{\prime \,
2}}}{2} \, \phi_c \hat{P} \right)^2 - \frac{1}{\xi_W} \abs{
\partial^{\mu} \hat{W}^{+}_{\mu} - i \, \xi_W \, \frac{g}{2} \, \phi_c
\hat{h}^{+} }^2 \, , \\
	\mathcal{L}_{\rm hd}^{\rm (gh.)} = 
	\bar{c}_A (-\partial^2) c_A 
	+ \bar{c}_Z \left( - \partial^2 - \xi_Z \frac{g^2 + g^{\prime \, 2}}{4} \phi_c^2 \right) c_Z 
	+ \bar{c}_{W^+} \left( - \partial^2 - \xi_W \frac{g^2}{4} \phi_c^2 \right) c_{W^+} \\
	\hspace{2cm}+ \bar{c}_{W^-} \left( - \partial^2 - \xi_W \frac{g^2}{4} \phi_c^2 \right) c_{W^-}
	+ {\rm interactions} \, ,
	\end{array} \nonumber
\end{align}
where 
\begin{align}
	&U_{\rm hd}(\hat{H}) = \lambda \left( \hat{H}^{\dagger}
\hat{H} - \frac{v^2}{2} \right)^2 \, , \\
	&\hat{D}_{\mu} H = \left( \partial_{\mu} - i g
\frac{\sigma^a}{2} \hat{W}_{\mu}^{a} - i g^{\prime} \frac{1}{2}
\hat{B}_{\mu} \right) H \, , \\
	& \hat{B}_{\mu \nu} = \partial_{\mu} \hat{B}_{\nu} -
\partial_{\nu} \hat{B}_{\mu} \, , \\
	& \hat{W}^{a}_{\mu \nu} = \partial_{\mu} \hat{W}^{a}_{\nu} -
\partial_{\nu} \hat{W}^{a}_{\mu} + g \epsilon^{abc} \hat{W}_{\mu}^{b}
\hat{W}_{\nu}^{c} \, , \\
	& (D^{\mu} \hat{W}_{\mu \nu})^{a} = \partial^{\mu} \hat{W}^{a}_{\mu \nu} + g \epsilon^{abc} \hat{W}^{b \, \mu} \hat{W}_{c \, \mu \nu}  \per
\end{align}
Since we are only interested in calculating the tree-level masses, we
drop the interactions (terms containing products of three or more
fields).  After expanding the Higgs field with \eref{eq:H_expansion}
and performing the rotation \eref{eq:ZAW_rotation}, the Lagrangian
becomes
\begin{align}
	&U_{\rm hd} = \frac{\lambda}{4} (\phi_c^2 - v^2)^2 + \lambda
\phi_c (\phi_c^2 - v^2) \hat{h} + \frac{1}{2} \lambda (3 \phi_c^2 -
v^2) \hat{h}^2 + \frac{1}{2} \lambda (\phi_c^2 - v^2) \hat{P}^2 +
\lambda (\phi_c^2 - v^2) \hat{h}^+ \hat{h}^- \, , \\
	&\mathcal{L}_{\rm hd}^{\rm (H)} + \mathcal{L}_{\rm hd}^{\rm (g.f.)} =
	\frac{1}{2} \left[ \left( \partial_{\mu} \hat{h} \right)^2 - \frac{1}{\LamH^2} (\partial^2 \hat{h})^2 \right]
	+ \frac{1}{2} \left[ \left( \partial_{\mu} \hat{P} \right)^2  - \frac{1}{\LamH^2} (\partial^2 \hat{P})^2 \right] \nn
	& \hspace{3cm}+ \left[ \abs{ \partial_{\mu} \hat{h}^+ }^2  - \frac{1}{\LamH^2} \abs{ \partial^2 \hat{h}^+ }^2 \right] \notag\\
	&\hspace{2.2cm}+ \frac{1}{2} \frac{g^2 + g^{\prime \, 2}}{4} \phi_c^2 \hat{Z}_{\mu} \hat{Z}^{\mu}
	+ \frac{g^2}{4} \phi_c^2 \abs{ \hat{W}_{\mu}^{-} }^2
	- \frac{1}{2} \frac{g^2 + g^{\prime \, 2}}{4} \xi_Z \phi_c^2 \hat{P}^2 
	- \xi_W \frac{g^2}{4} \phi_c^2 \abs{ \hat{h}^{+} }^2 \notag\\
	&\hspace{2.2cm}- \frac{1}{2 \xi_A} (\partial_{\mu} \hat{A}^{\mu} )^2
	- \frac{1}{\xi_W} \abs{ \partial_{\mu} \hat{W}^{- \, \mu} }^2 
	- \frac{1}{2 \xi_Z} (\partial_{\mu} \hat{Z}^{\mu} )^2 \notag\\
	\label{eq:tot_deriv_terms}
	&\hspace{2.2cm}+ \frac{ \sqrt{g^2 + g^{\prime \, 2}}}{2} \phi_c \partial_{\mu} \left( \hat{P} \hat{Z}^{\mu} \right)
	+ \frac{g}{2} \phi_c \partial_{\mu} \left( i \, \hat{h}^+
\hat{W}^{- \mu} - i \, \hat{h}^{-} \hat{W}^{+ \, \mu} \right) \, , \\
	&\mathcal{L}_{\rm hd}^{\rm (B)} + \mathcal{L}_{\rm hd}^{\rm (W)} =
	- \frac{1}{4} \left( \partial_{\mu} \hat{A}_{\nu} - \partial_{\nu} \hat{A}_{\mu} \right)^2
	- \frac{1}{4} \left( \partial_{\mu} \hat{Z}_{\nu} - \partial_{\nu} \hat{Z}_{\mu} \right)^2
	- \frac{1}{2} \abs{ \partial_{\mu} \hat{W}^{-}_{\nu} - \partial_{\nu} \hat{W}^{-}_{\mu} }^2 \notag\\
	&\hspace{2.2cm}+ \frac{1}{\LamW^2} \abs{ \partial^2 \hat{W}_{\mu}^{-} - \partial_{\mu} \partial^{\nu} \hat{W}_{\nu}^{-} }^2 
	+ \frac{1}{2 \LamZ^2} \left( \partial^2 \hat{Z}_{\mu} - \partial_{\mu} \partial^{\nu} \hat{Z}_{\nu} \right)^2 \notag\\
	&\hspace{2.2cm}+ \frac{1}{2 \LamA^2} \left( \partial^2 \hat{A}_{\mu} - \partial_{\mu} \partial^{\nu} \hat{A}_{\nu} \right)^2 
	- \frac{1}{2 \LamAZ^2} \left( \partial^2 \hat{A}_{\mu} -
\partial_{\mu} \partial^{\nu} \hat{A}_{\nu} \right) \left( \partial^2
\hat{Z}^{\mu} \! - \partial^{\mu} \partial^{\alpha} \hat{Z}_{\alpha}
\right) ,
\end{align}
where we have defined
\begin{align}\label{eq:LamAZ_defs}
\begin{array}{l}
	\LamA \equiv \left( \frac{\cos^2 \theta_W}{\LamB^2} +
\frac{\sin^2 \theta_W}{\LamW^2} \right)^{-1/2} \, , \\
	\LamZ \equiv \left( \frac{\sin^2 \theta_W}{\LamB^2} +
\frac{\cos^2 \theta_W}{\LamW^2} \right)^{-1/2} \, , \\
	\LamAZ \equiv \left( \frac{\sin 2 \theta_W}{\LamB^2} -
\frac{\sin 2 \theta_W}{\LamW^2} \right)^{-1/2} \, .
\end{array}
\end{align}
The final two terms in \eref{eq:tot_deriv_terms} are total derivatives and
can be dropped.  After integrating by parts and dropping total
derivative terms, one obtains
\begin{align}
	\mathcal{L}_{\rm hd}^{\rm (EW)} =& 
	- \frac{\lambda}{4} (\phi_c^2 - v^2)^2 
	- \lambda \phi_c (\phi_c^2 - v^2) \hat{h} \label{eq:LEW_linearterm}\\
	&+ \frac{1}{2} \hat{h} \left( -\partial^2 - \frac{1}{\LamH^2} \partial^4 - {m}^2_{\hat{h}}\right) \hat{h} 
	+ \frac{1}{2} \hat{P} \left( - \partial^2 - \frac{1}{\LamH^2} \partial^4 - {m}^2_{\hat{P}}\right) \hat{P} \nn
	& \hspace{1cm} + \hat{h}^{+} \left( - \partial^2 - \frac{1}{\LamH^2} \partial^4 - {m}^2_{\hat{h}^{\pm}} \right) \hat{h}^{-} \notag\\
	&+ \frac{1}{2} \hat{A}^{\mu} \left[ -g_{\mu \nu} \left( -\partial^2 - \frac{\partial^4}{\LamA^2} - {m}^2_{\hat{A}} \right) + \left( -\frac{\partial^2}{\LamA^2} - 1 + \frac{1}{\xi_A} \right) \partial_{\mu} \partial_{\nu} \right] \hat{A}^{\nu} \notag\\
	&+ \frac{1}{2} \hat{Z}^{\mu} \left[ -g_{\mu \nu} \left( -\partial^2 - \frac{\partial^4}{\LamZ^2} - {m}^2_{\hat{Z}} \right) + \left( -\frac{\partial^2}{\LamZ^2} - 1 + \frac{1}{\xi_Z} \right) \partial_{\mu} \partial_{\nu} \right] \hat{Z}^{\nu} \notag\\
	&+\frac{1}{2} \hat{A}^{\mu} \left[ - \left( g_{\mu \nu} \partial^2 - \partial_{\mu} \partial_{\nu} \right) \frac{\partial^2 }{\LamAZ^2} \right] \hat{Z}^{\nu} \label{eq:AZ_mixing} \notag\\
	&+ \hat{W}^{+ \, \mu} \left[ -g_{\mu \nu} \left( -\partial^2 - \frac{\partial^4}{\LamW^2} - {m}^2_{\hat{W}^{\pm}} \right) + \left(-  \frac{\partial^2}{\LamW^2} - 1 + \frac{1}{\xi_W} \right) \partial_{\mu} \partial_{\nu} \right] \hat{W}^{- \, \nu} \notag\\
	&+ \bar{c}_A (-\partial^2) c_A 
	+ \bar{c}_Z \left( - \partial^2 - \xi_Z {m}_{\hat{Z}}^2 \right) c_Z
	+ \bar{c}_{W^+} \left( - \partial^2 - \xi_W {m}_{\hat{W}^{\pm}}^2 \right) c_{W^+} \nn
	&\hspace{1cm} + \bar{c}_{W^-} \left( - \partial^2 - \xi_W
{m}_{\hat{W}^{\pm}}^2 \right) c_{W^-} \, , 
\end{align}
where
\begin{align}
\begin{array}{lll}
	{m}_{\hat{h}}^2 &\equiv& \lambda (3 \phi_c^2 - v^2) \, , \\
	{m}_{\hat{P}}^2 &\equiv& \lambda (\phi_c^2 - v^2) + \xi_{Z}
{m}_{\hat{Z}}^2 \, , \\
	{m}_{\hat{h}^{\pm}}^2 &\equiv& \lambda (\phi_c^2 - v^2) +
\xi_{W} {m}_{\hat{Z}}^2 \, ,
\end{array}
\hspace{2cm}
\begin{array}{lll}
	{m}_{\hat{W}^{\pm}}^2 &\equiv& \frac{g^2}{4} \phi_c^2 \, , \\
	{m}_{\hat{Z}}^2 &\equiv& \frac{g^2+g^{\prime \, 2}}{4}
\phi_c^2 \, , \\
	{m}_{\hat{A}}^2 &\equiv& 0 \, . \\
\end{array}
\end{align}
With the Lagrangian in this form, it is straightforward to read off
the propagators.  For the scalars one finds
\begin{align}\label{eq:scalar_props}
	& D_{\hat{h}}(p) = i \left( p^2 - \frac{p^4}{\LamH^2} - {m}_{\hat{h}}^2 \right)^{-1} 
	= \frac{\LamH^2}{{m}_{\tilde{h}}^2 - {m}_{h}^2} \left(
\frac{i}{p^2 - {m}_h^2} - \frac{i}{p^2 - {m}_{\tilde{h}}^2} \right) \,
, \nn
	& D_{\hat{P}}(p) = i \left( p^2 - \frac{p^4}{\LamH^2} - {m}_{\hat{P}}^2 \right)^{-1} 
	= \frac{\LamH^2}{{m}_{\tilde{P}}^2 - {m}_{P}^2} \left(
\frac{i}{p^2 - {m}_P^2} - \frac{i}{p^2 - {m}_{\tilde{P}}^2} \right) \,
, \nn
	& D_{\hat{h}^{\pm}}(p) = i \left( p^2 - \frac{p^4}{\LamH^2} - {m}_{\hat{h}^{\pm}}^2 \right)^{-1} 
	= \frac{\LamH^2}{{m}_{\tilde{h}^{\pm}}^2 - {m}_{h^{\pm}}^2}
\left( \frac{i}{p^2 - {m}_{h^{\pm}}^2} - \frac{i}{p^2 -
{m}_{\tilde{h}^{\pm}}^2} \right) \, ,
\end{align}
where
\begin{align}\label{eq:scalar_masses}
\begin{array}{lll}
	&&\text{SM-like Pole} \\
	{m}_h^2 &=& \frac{\LamH^2}{2} \left( 1 - \sqrt{1 - \frac{4
{m}_{\hat{h}}^2}{\LamH^2}} \right) \, , \\
	{m}_P^2 &=& \frac{\LamH^2}{2} \left( 1 - \sqrt{1 - \frac{4
{m}_{\hat{P}}^2}{\LamH^2}} \right) \, , \\
	{m}_{h^{\pm}}^2 &=& \frac{\LamH^2}{2} \left( 1 - \sqrt{1 -
\frac{4 {m}_{\hat{h}^{\pm}}^2}{\LamH^2}} \right) \, ,
\end{array}
\hspace{2cm}
\begin{array}{lll}
	&&\text{LW-like Pole} \\
	{m}_{\tilde{h}}^2 &=& \frac{\LamH^2}{2} \left( 1 + \sqrt{1 -
\frac{4 {m}_{\hat{h}}^2}{\LamH^2}} \right) \, , \\
	{m}_{\tilde{P}}^2 &=& \frac{\LamH^2}{2} \left( 1 + \sqrt{1 -
\frac{4 {m}_{\hat{P}}^2}{\LamH^2}} \right) \, , \\
	{m}_{\tilde{h}^{\pm}}^2 &=& \frac{\LamH^2}{2} \left( 1 +
\sqrt{1 - \frac{4 {m}_{\hat{h}^{\pm}}^2}{\LamH^2}} \right) \, .
\end{array}
\end{align}
The poles are classified as ``SM-like'' or ``LW-like'', depending on
whether the residue of the pole is positive or negative.

In the gauge sector, the ghost propagators are immediately seen to be
\begin{align}\label{eq:ghost_props}
\begin{array}{l}
	D_{c_A}(p) = \frac{i}{p^2} \, , \\
	D_{c_Z}(p) = \frac{i}{p^2 - \xi_Z \, m_{\hat{Z}}^2} \, , \\
	D_{c_{W^+}}(p) = \frac{i}{p^2 - \xi_W \, m_{\hat{W}^{\pm}}^2}
\, , \\
	D_{c_{W^-}}(p) = \frac{i}{p^2 - \xi_W \, m_{\hat{W}^{\pm}}^2}
\, .
\end{array}
\end{align}
We define the transverse and longitudinal projection operators
$\Pi_T^{\mu \nu} \equiv g^{\mu \nu} - p^{\mu} p^{\nu} / p^2$ and
$\Pi_L^{\mu \nu} \equiv p^{\mu} p^{\nu} / p^2$, and obtain
\begin{align}\label{eq:W_boson_prop}
	D^{\mu \nu}_{\hat{W}^{\pm}}(p) =&
	-i \, \Pi_T^{\mu \nu}(p) \left( p^2 - \frac{p^4}{\LamW^2} - {m}_{\hat{W}^{\pm}}^2 \right)^{-1}
	- i \, \Pi_L^{\mu \nu}(p) \left( \frac{p^2}{\xi_W} -
{m}_{\hat{W}^{\pm}}^2 \right)^{-1} \nn
	=& \frac{\LamW^2}{{m}_{\tilde{W}^{\pm}}^2 - {m}_{W^{\pm}}^2} \left( \frac{-i \, \Pi^{\mu \nu}_{T}(p) }{p^2 - {m}_{W^{\pm}}^2} - \frac{-i \, \Pi^{\mu \nu}_{T}(p) }{p^2 - {m}_{\tilde{W}^{\pm}}^2} \right) 
	+ \frac{-i \, \xi_W \, \Pi^{\mu \nu}_{L}(p) }{p^2 - \xi_W
{m}_{\hat{W}^{\pm}}^2} \, ,
\end{align}
where
\begin{align}\label{eq:gauge_masses}
\begin{array}{lll}
	&&\text{SM-like Pole} \\
	{m}_{W^{\pm}}^2 &=& \frac{\LamW^2}{2} \left( 1 - \sqrt{1 -
\frac{4 {m}_{\hat{W}^{\pm}}^2}{\LamW^2}} \right) \, ,
\end{array}
\hspace{2cm}
\begin{array}{lll}
	&&\text{LW-like Pole} \\
	{m}_{\tilde{W}^{\pm}}^2 &=& \frac{\LamW^2}{2} \left( 1 +
\sqrt{1 - \frac{4 {m}_{\hat{W}^{\pm}}^2}{\LamW^2}} \right) \, .
\end{array}
\end{align}
We defer a discussion of the longitudinal polarization state
until the end.  The term on line \eref{eq:AZ_mixing} corresponds to a
mixing between transverse polarizations of $\hat{A}^{\mu}$ and
$\hat{Z}^{\mu}$, which gives rise to off-diagonal terms in the inverse
propagator:
\begin{align}
	(D_{\hat{A} \hat{Z}}^{-1})^{\mu \nu}(p) = 
	i \, \Pi_{T}^{\mu \nu} \begin{pmatrix}
	p^2 - \frac{p^4}{\LamA^2} - {m}_{\hat{A}}^2 & \frac{p^4}{2 \LamAZ^2} \\ \frac{p^4}{2 \LamAZ^2} & p^2 - \frac{p^4}{\LamZ^2} - {m}_{\hat{Z}}^2
	\end{pmatrix}
	+ i \, \Pi_{L}^{\mu \nu} \begin{pmatrix}
	\frac{p^2}{\xi_A} - {m}_{\hat{A}}^2 & 0 \\ 0 & \frac{p^2}{\xi_Z} - {m}_{\hat{Z}}^2
	\end{pmatrix} \per
\end{align}
For simplicity, we assume just one common LW scale in the EW gauge
sector.  Then one has $\LamB = \LamW = \LamA = \LamZ \equiv \LamEW$
and also $(\LamAZ)^{-2} = 0$ using \eref{eq:LamAZ_defs}.  The mixing
vanishes and the propagators become
\begin{align}
	D^{\mu \nu}_{\hat{A}}(p) =& \frac{\LamEW^2}{{m}_{\tilde{A}}^2
- {m}_{A}^2} \left( \frac{-i \, \Pi^{\mu \nu}_{T}(p) }{p^2 -
{m}_{A}^2} - \frac{-i \, \Pi^{\mu \nu}_{T}(p) }{p^2 -
{m}_{\tilde{A}}^2} \right) + \frac{-i \, \xi_A \, \Pi^{\mu \nu}_{L}(p)
}{p^2 - \xi_A {m}_{\hat{A}}^2} \label{eq:photon_prop} \, , \\
	D^{\mu \nu}_{\hat{Z}}(p) =& \frac{\LamEW^2}{{m}_{\tilde{Z}}^2
- {m}_{Z}^2} \left( \frac{-i \, \Pi^{\mu \nu}_{T}(p) }{p^2 -
{m}_{Z}^2} - \frac{-i \, \Pi^{\mu \nu}_{T}(p) }{p^2 -
{m}_{\tilde{Z}}^2} \right) + \frac{-i \, \xi_Z \, \Pi^{\mu \nu}_{L}(p)
}{p^2 - \xi_Z {m}_{\hat{Z}}^2} \, ,
\end{align}
where
\begin{align}\label{eq:gauge_masses2}
\begin{array}{lll}
	&&\text{SM-like Pole} \\
	{m}_{A}^2 &=& \frac{\LamEW^2}{2} \left( 1 - \sqrt{1 - \frac{4
{m}_{\hat{A}}^2}{\LamEW^2}} \right) = 0 \, , \\
	{m}_{Z}^2 &=& \frac{\LamEW^2}{2} \left( 1 - \sqrt{1 - \frac{4
{m}_{\hat{Z}}^2}{\LamEW^2}} \right) \, ,
\end{array}
\hspace{1.5cm}
\begin{array}{lll}
	&&\text{LW-like Pole} \\
	{m}_{\tilde{A}}^2 &=& \frac{\LamEW^2}{2} \left( 1 + \sqrt{1 -
\frac{4 {m}_{\hat{A}}^2}{\LamEW^2}} \right) = \LamEW^2 \, , \\
	{m}_{\tilde{Z}}^2 &=& \frac{\LamEW^2}{2} \left( 1 + \sqrt{1 -
\frac{4 {m}_{\hat{Z}}^2}{\LamEW^2}} \right) \, .
\end{array}
\end{align}
Note that the photon is massless, and that the mass of its LW partner
is independent of $\phi_c$.

Having calculated the spectrum, let us discuss the counting of degrees
of freedom.  The scalar propagators \eref{eq:scalar_props} reveal that
each of the fields $\hat{h}, \hat{P}, \hat{h}^+,$ and $\hat{h}^-$
carries two degrees of freedom: a lighter SM-like resonance and a
heavier LW-like resonance.  We might expect this doubling to carry
over to the gauge fields as well, but an inspection of their
propagators reveals that this is not the case.  In counting the gauge
boson degrees of freedom, note that ${\rm Tr} \, \Pi_T = \Pi_{T , \,
\mu \nu} g^{\mu \nu} = 3$ and ${\rm Tr} \, \Pi_L = 1$.  Examining the
propagator \eref{eq:photon_prop}, we see that the $\hat{A}$ contains
seven degrees of freedom: three massless transverse polarizations
(${m}_A^2 = 0$), one massless longitudinal polarization
(${m}_{\hat{A}}^2 = 0$), and three massive transverse polarizations
(${m}_{\tilde{A}}^2 = \LamEW^2$).  The four massless degrees of
freedom constitute the SM photon, and after accounting for the two
``negative degrees of freedom'' of the ghosts $c_A$ and $\bar{c}_A$,
the count of ``physical'' photon polarizations is reduced to two.
Here, the LWSM does not double the number of gauge degrees of freedom,
but instead adds three, which is what one expects for an additional
massive resonance.  For the $\hat{Z}$ boson we count three degrees of
freedom with mass ${m}_Z^2$, three degrees of freedom with mass
${m}_{\tilde{Z}}^2$, one degree of freedom with mass $\xi_Z \,
{m}_{\hat{Z}}^2$, and two negative degrees of freedom of mass $\xi_Z
\, {m}_{\hat{Z}}^2$ coming from the ghosts.  The ghost cancels the
longitudinal polarization state, and one negative degree of freedom
remains.  Once we restrict to the Landau gauge ($\xi_A = \xi_Z = \xi_W
= 0$), the ghosts and longitudinal polarizations become massless.
Then these degrees of freedom do not yield a field-dependent
contribution to the effective potential, but they do affect the number
of relativistic species at finite temperature.  Thus, we have counted
them as massless particles in \tref{tab:LWSM_masses}, which also
summarizes Eqs.~(\ref{eq:scalar_masses}), (\ref{eq:gauge_masses}), and
(\ref{eq:gauge_masses2}).

\subsection{Top Sector}\label{sub:LWSM_fermion}

Let the $\SU{2}$ doublet $\hat{Q}_L = (\hat{u}_L \, , \, \hat{d}_L
)^T$ be a left-handed Weyl spinor, and let the singlet $\hat{u}_R$ be
a right-handed Weyl spinor.  Neglecting gauge interactions, the
Lagrangian for the top sector is written as
\begin{align}
	\mathcal{L}_{\rm hd}^{\rm (top)} = \, & 
	(\hat{Q}_L)^{\dagger} i \bar{\slashed{\partial}} \hat{Q}_L
	+ \frac{1}{\LamQ^2} (\hat{Q}_L)^{\dagger} i \bar{\slashed{\partial}} \slashed{\partial} \bar{\slashed{\partial}} \hat{Q}_L
	+ (\hat{u}_R)^{\dagger} i \slashed{\partial} \hat{u}_R 
	+ \frac{1}{\Lamu^2} (\hat{u}_R)^{\dagger} i \slashed{\partial}
\bar{\slashed{\partial}} \slashed{\partial} \hat{u}_R \, , \nn
	&- h_t \left( (\hat{Q}_L)^{\dagger} {\bm \epsilon}
\hat{H}^{\ast} \hat{u}_R - (\hat{u}_R)^{\dagger} \hat{H} {\bm
\epsilon} \hat{Q}_L \right) \, ,
\end{align}
where $\slashed{\partial} = \sigma^{\mu} \partial_{\mu}$ and
$\bar{\slashed{\partial}} = \bar{\sigma}^{\mu} \partial_{\mu}$.
Contractions of the $\SU{2}$ doublets is accomplished with the totally
antisymmetric 2-tensor ${\bm \epsilon}$.  After electroweak symmetry
breaking, one replaces $\hat{H} \to ( 0 \, , \, \phi_c / \sqrt{2}
)^T$, and obtains
\begin{align}
	\mathcal{L}_{\rm hd}^{\rm (top)} =  \, &
	(\hat{u}_L)^{\dagger} \left( i \bar{\slashed{\partial}} + \frac{i \bar{\slashed{\partial}} \slashed{\partial} \bar{\slashed{\partial}}}{\LamQ^2} \right) \hat{u}_L
	+ (\hat{u}_R)^{\dagger} \left( i \slashed{\partial} + \frac{i \slashed{\partial} \bar{\slashed{\partial}} \slashed{\partial}}{\Lamu^2} \right) \hat{u}_R
	-\frac{h_t \, \phi_c}{\sqrt{2}} \left[ (\hat{u}_L)^{\dagger}  \hat{u}_R + (\hat{u}_R)^{\dagger} \hat{u}_L \right] \per
\end{align}
One can now collect the Weyl spinors into the Dirac spinor $\hat{t} =
( \, \hat{u}_L \, , \, \hat{u}_R \, )^T$.  Using the standard
definitions
\begin{align}
	\gamma^{\mu} = \left( \begin{smallmatrix} 0 & \sigma^{\mu} \\ \bar{\sigma}^{\mu} & 0 \end{smallmatrix} \right) 
	\hspace{0.4cm} , \hspace{0.4cm} 
	\bar{\hat{t}} \equiv \hat{t}^{\dagger} \gamma^0 
	\hspace{0.4cm} , \hspace{0.4cm} 
	\slashed{\partial} \hat{t} = \gamma^{\mu} \partial_{\mu} \hat{t} 
	\hspace{0.4cm} , \hspace{0.4cm} 
	\gamma^5 = i \gamma^0 \gamma^1 \gamma^2 \gamma^3 
	\hspace{0.4cm} , \hspace{0.4cm} 
	P_{L,R} = \frac{1 \mp \gamma^5}{2} \nonumber \com
\end{align}
the Lagrangian can be written as 
\begin{align}
	\mathcal{L}_{\rm hd}^{\rm (top)} = 
	\bar{\hat{t}} \left( i \slashed{\partial} + \frac{i \slashed{\partial}^3}{\LamQ^2} P_L + \frac{i \slashed{\partial}^3}{\Lamu^2} P_R \right) \hat{t}
	- \bar{\hat{t}} {m}_{\hat{t}} \hat{t} \, ,
\end{align}
where ${m}_{\hat{t}} \equiv h_t \phi_c / \sqrt{2}$.  To simplify, we
assume that $\LamQ = \Lamu \equiv \Lamt$.  Then the Lagrangian reduces to
\eref{eq:LWfermiontoy_L}, and the propagator is
\begin{align}
	D_{\hat{t}}(p) = \ & i \left( -\frac{\slashed{p}^3}{\Lamt^2} +
\slashed{p} - {m}_{\hat{t}}(\phi_c) \right)^{-1} \nn
	= \ & + \frac{\Lamt^2}{({m}_{\tilde{t}_1} - {m}_{t}) ({m}_{t}
- {m}_{\tilde{t}_2})} \frac{i}{\slashed{p} - {m}_{t}} \nn
	&- \frac{\Lamt^2}{({m}_{\tilde{t}_1} - {m}_{t})
({m}_{\tilde{t}_1} - {m}_{\tilde{t}_2})} \frac{i}{\slashed{p} -
{m}_{\tilde{t}_1}} \nn
	&- \frac{\Lamt^2}{({m}_{t} - {m}_{\tilde{t}_2})
({m}_{\tilde{t}_1} - {m}_{\tilde{t}_2})} \frac{i}{\slashed{p} -
{m}_{\tilde{t}_2}} \, ,
\end{align}
where 
\begin{align}
\begin{array}{lcl}
	\text{SM-like Pole:} & \qquad & {m}_{t}(\phi_c) \equiv \Lamt
\sqrt{\frac{2}{3} \left( 1 - \cos \frac{\theta_t}{3} \right)} \, , \\
	\text{LW-like Pole:} & \qquad & {m}_{\tilde{t}_1}(\phi_c)
\equiv \Lamt \sqrt{ \frac{2}{3} \left( 1 + \cos \frac{\theta_t +
\pi}{3} \right) } \, , \\
	\text{LW-like Pole:} & \qquad & {m}_{\tilde{t}_2}(\phi_c)
\equiv - \Lamt \sqrt{ \frac{2}{3} \left( 1 + \cos \frac{\theta_t -
\pi}{3} \right) } \, ,
\end{array} 
\end{align}
where $\theta_t \equiv \arctan \, \frac{2 \sqrt{\alpha
    (1-\alpha)}}{1-2 \alpha}$ and $\alpha \equiv \frac{27}{4}
\frac{{m}_{\hat{t}}^2}{\Lamt^2}$.  The angle $0 \leq \theta_t \leq
\pi$ is in the first or second quadrant, and the LW stability
condition imposes $\alpha < 1$.

\end{appendix}

\bibliographystyle{JHEP}

\providecommand{\href}[2]{#2}\begingroup\raggedright\endgroup

\end{document}